\documentclass[aps,prb,twocolumn,floatfix]{revtex4}
\usepackage{graphicx}
\usepackage{epstopdf}
\usepackage{amssymb}
\usepackage{epsf}
\newcommand{\bra}[1]{\langle #1|}
\newcommand{\ket}[1]{|#1\rangle}
\newcommand{\braket}[1]{\left\langle#1\right\rangle}

\renewcommand{\phi}{\varphi}
\renewcommand{\epsilon}{\varepsilon}

\renewcommand{\vec}[1]{{\bf #1}}

\newcommand{\eqnref}[1]{Eq.~(\ref{#1})}	
\newcommand{\fref}[1]{Fig.~\ref{#1}}	
\newcommand{\secref}[1]{Section~\ref{#1}}

\newcommand{\beq}{\begin{equation}}
\newcommand{\eeq}{\end{equation}}
\newcommand{\ba}{\begin{array}{ccc}}
\newcommand{\ea}{\end{array}}

\usepackage{amssymb}
\usepackage{amsmath}

\begin{document}
\title{Topological phenomena in quantum walks; \\ elementary introduction to the physics of topological phases}
\author{Takuya Kitagawa} 
\affiliation{Physics Department, Harvard University, Cambridge,
Massachusetts 02138, USA}

\date{\today}
\begin{abstract}
Discrete quantum walks are dynamical protocols for controlling a single quantum particle. 
Despite of its simplicity, quantum walks display rich topological phenomena and provide 
one of the simplest systems to study and understand topological phases. 
In this article, we review the physics of discrete quantum walks in one and two dimensions in light of 
topological phenomena
and provide  elementary explanations of topological phases and their physical consequence, namely
the existence of boundary states. 
We demonstrate that quantum walks are versatile systems that simulate many topological phases 
whose classifications are known for static Hamiltonians. 
Furthermore, topological phenomena appearing in quantum walks go beyond what has been known
in static systems; there are phenomena unique to quantum walks, being an example of periodically driven systems, 
that do not exist in static systems. 
Thus the quantum walks not only provide a powerful tool as a quantum simulator for static topological phases 
but also give unique opportunity to study topological phenomena in driven systems. 
\end{abstract}

\maketitle 
 \section{Introduction}
\begin{figure}[t]
\begin{center}
\includegraphics[width = 6cm]{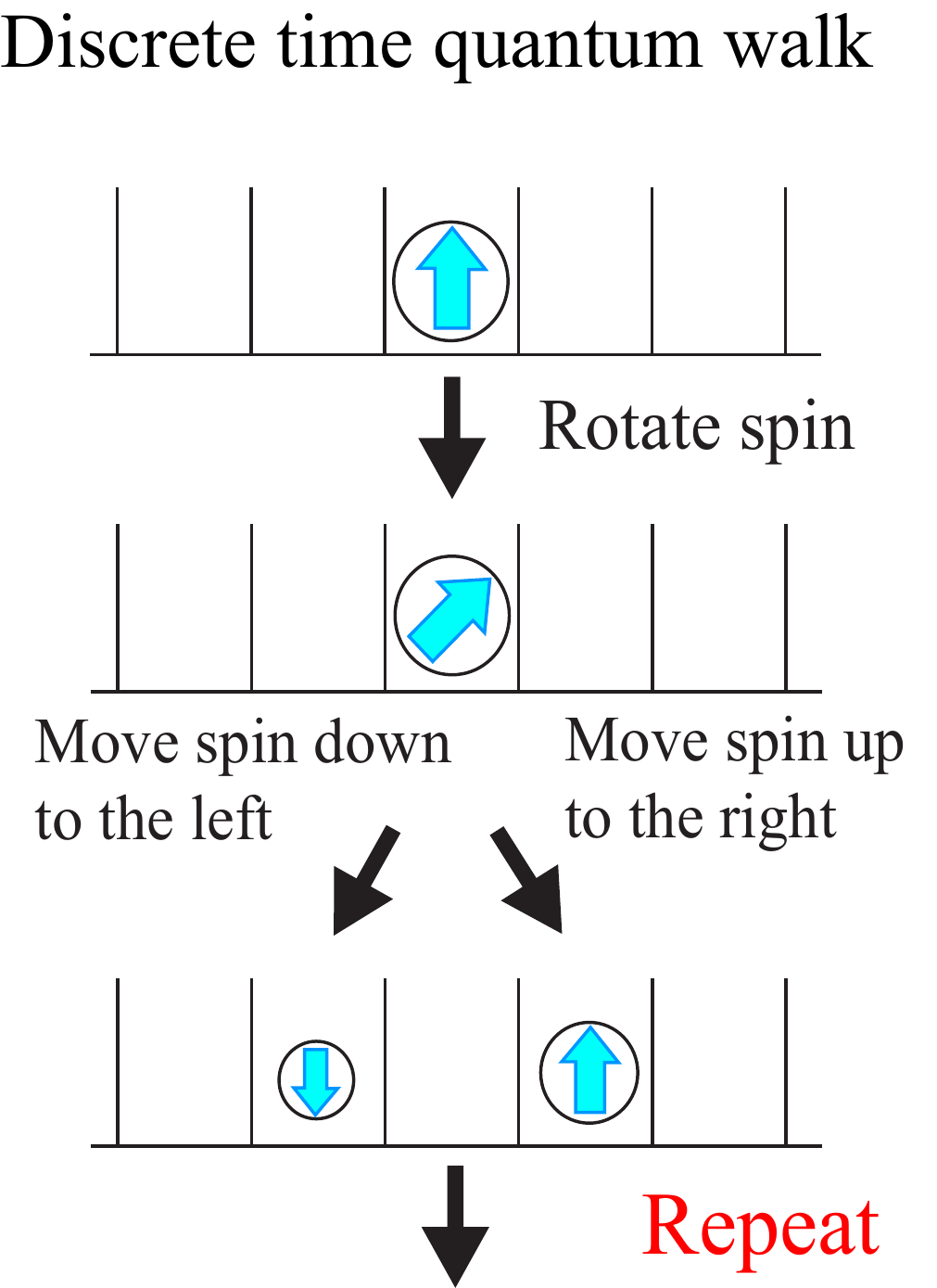}
\caption{The protocol of a conventional discrete quantum walk. 
A conventional quantum walk is a dynamical protocol for controlling a spin $1/2$ quantum particle
in one dimensional lattice. It consists of two operations; 1. rotation around $y$ axis by angle $\theta$, whose 
operator is given by $R_{y}(\theta) = e^{-i\theta \sigma_{y}/2 }$; 2. spin-dependent translation $T$ where spin up particle
is move to the right by one lattice site and spin down particle is moved to the left by one lattice site. 
One step of the quantum walk is given by $U = T R_{y}(\theta)$ and the evolution of the particle after many steps
are studied.  }
\label{fig:conv_qw}
\end{center}
\end{figure}

 Discrete quantum walk, in its simplest form, is a dynamical protocol for controlling a single spin $1/2$ particle
 in one dimensional lattice (see \fref{fig:conv_qw}). 
 It consists of two operations given by a spin rotation and spin-dependent translation. 
 The evolution of quantum walk results from the repeated applications of these two operations in alternate fashion. 
From the first introduction of the concept of quantum walks by Aharonov\cite{Aharonov1993}, 
quantum walks attracted tremendous attentions 
due to their implications in quantum information science\cite{Kempe2003}. 
One of the attractive features of quantum walks is 
its simplicity which allows any student of physics who has a basic understanding of quantum mechanics to grasp 
its definition. Yet, the consequence of quantum walks is profound; on one hand, it provides a powerful tool for 
quantum algorithms\cite{Kempe2003, Farhi1998} and on the other, it displays rich topological phenomena revealing the deep relation between physics and the abstract field of mathematics\cite{Kitagawa2010}. 
In this article, we review the topological phenomena appearing in discrete quantum walks. 
While the concept of topological phases is often challenging to understand because it tends to be intimately 
intertwined with the physics of solid state materials, discrete quantum walks provide a rare opportunity to explain the idea of topological phases in the most elementary form due to its simplicity. 

This review article is organized as follows.
In \secref{sec:1DQW}, we study the physics of a one dimensional quantum walk. 
First, in \secref{sec:intro}, we define the simplest one dimensional quantum walk and give the description of 
quantum walks through so-called effective Hamiltonians. 
This quantum walk possesses a symmetry that is not apparent at a first sight but plays a crucial role for
its topological properties, and we describe this symmetry in \secref{sec:symmetry}. 
The effective Hamiltonian approach to quantum walks gives an intuition behind the behavior of quantum walks,
which we use to derive the analytic expression for their asymptotic distribution in \secref{sec:asymptotic}. 
This conventional quantum walk and slight variations of it has been realized in a numerous 
experiments with different physical settings, and we explain some of their realizations 
with cold atoms, photons and ions in \secref{sec:experiment}. 

In \secref{sec:topological_phases}, we first briefly review the main ideas of topological phases
that will be relevant to the study of quantum walks. 

Starting from \secref{section:topology}, we investigate the 
topological nature of the quantum walks. In \secref{sec:winding}, 
we describe the concept of topological invariants in the context of 
quantum walks. In order to fully explore the topological phases in quantum walks,
we extend the conventional quantum walks to so-called 
split-step quantum walk in \secref{sec:splitstep}, which displays multiple topological phases in its phase diagram.
Using split-step quantum walks, we argue in \secref{sec:bound_state} that 
physical manifestations of topological nature of quantum walks are the appearance of 
bound states across distinct topological phases.  We demonstrate the existence of such bound states
in two physically distinct situations; one is inhomogeneous quantum walks where rotation angles are 
varied in space (\secref{sec:bound_state})
and the other is the quantum walks with reflecting boundary (\secref{sec:reflecting}). 
The unique nature of these bound states lies in the robustness of their existence against small perturbations. 
We provide additional understandings of this robustness in \secref{sec:topological_invariant}, 
which is based on the spectrum and topological invariants 
associated with the bound states.
These bound states in the quantum walk have the same topological origin as the bound states predicted to arise 
in polyacetylene described by the Su-Schrieffer-Heeger model and Jackiw-Rebbi model in high energy physics. 

Remarkably, there are also phenomena in quantum walks that have not been predicted before in static systems; 
the existence
of two flavors of bound states at quasi-energy $E=0$ and $E=\pi$. This phenomenon is unique to periodically
driven systems, and we illustrate the physics in \secref{sec:zeropi}. These one dimensional
topological phenomena in quantum walks have been experimentally verified in [\onlinecite{Kitagawa2011}].

From \secref{sec:2DQW}, we extend the idea of quantum walks to two dimension and study their 
topological properties. In \secref{sec:2D_quantum_walk}, we define the two dimensional quantum walks that 
display non-zero Chern numbers, 
which is the topological invariant responsible for integer quantum Hall effects. We explain in detail
how Chern number arises in quantum walks in this section and demonstrate that non-trivial topology in this class results 
in unidirectionally propagating modes at the edge of the systems. 
As is the case for one dimensional quantum walks, two dimensional quantum walks display topological phenomena
unique to periodically driven systems. In \secref{sec: simple_2D}, we describe a simple quantum walk which 
possesses unidirectionally propagating modes as a result of topological invariants unique to driven systems,
namely the non-trivial winding number in energy direction. 
Due to its simplicity, this quantum walk has the advantage of being easier to implement in experiments compared
to the quantum walk introduced in \secref{sec:2D_quantum_walk}. 

In \secref{sec:others}, we briefly discuss the realization of other topological phases, using quantum walks. 
In \secref{sec:conclusion}, we conclude with possible extensions of ideas reviewed in this article.

\section{One dimensional quantum walk : general property} \label{sec:1DQW}
\subsection{Effective Hamiltonian description of quantum walk} \label{sec:intro}
The simplest form of discrete quantum walks\cite{dqw} is defined as a protocol acting on a single particle 
in one dimensional lattice with two internal degrees of freedom. In an analogy with spin $1/2$ particle, 
we refer to the internal degrees of freedom as as "spin up" 
and "spin down." This quantum walk, which we call as a conventional quantum walk in this article,  
consists of two operations(see \fref{fig:conv_qw}); 
\begin{enumerate}
\item rotation of the spin around $y$ axis by
angle $\theta$, corresponding to the operation $R_{y}(\theta)= e^{-i\theta \sigma_{y}/2}$ where $\sigma_{y}$ 
is a Pauli operator. 
\item spin-dependent translation $T$ of the particle, where spin up particle
is move to the right by one lattice site and spin down particle is moved to the left by one lattice site..
Explicitly, $T = \sum_{j} \ket{j+1} \bra{j} \otimes \ket{\uparrow} \bra{\uparrow} 
+  \ket{j-1} \bra{j} \otimes \ket{\downarrow} \bra{\downarrow}$.  
\end{enumerate}
These two operations make up one step $U= TR_{y}(\theta)$ of the quantum walk, and the evolution of the particle 
after many steps of the walk is studied. 
It is possible to define more general quantum walks by replacing the first operation by 
any unitary operation $R(\theta, \phi)$, which can be written as the product of 
the rotation of the spin around some axis $\vec{n}$ by angle $\theta$ and the phase accumulation $\phi$
{\it i.e.} $R(\theta, \phi) = e^{-i\theta \vec{n} \cdot \vec{\sigma}/2} e^{-i\phi}$. However, many central concepts 
of topological phases in quantum walks
can be illustrated with the simple quantum walk defined above, so we focus on the conventional quantum walk
in this paper. The extensions to more general case is straightforward. Discrete quantum walks have been realized in variety of experiments with ions, cold atoms and photons
\cite{Zhringer2010, Karsi2009, Schreiber2010, Broome2010}. 
In \secref{sec:experiment}, we describe experimental realizations of quantum walks in some of these systems 
in details. 
 
Many properties of quantum walks, such as the distribution of the particle after many steps, have been extensively
studied from mathematical physics point of view\cite{Konno, Grimmett}. In this paper, we take the intuitive picture in which 
quantum walk is considered as a stroboscopic realization of static effective Hamiltonian, defined through
the evolution operator of one step quantum walk protocol $U= TR_{y}(\theta)$;
\begin{eqnarray}
U &=& TR_{y}(\theta)\\ 
 &\equiv& e^{-iH_{\textrm{eff}}T} \label{effective_Hamiltonian}
\end{eqnarray} 
Here, $T$ is the time it takes to carry out one step of the quantum walk. Because $n$ steps of quantum walk 
correspond to the evolution $U^{n} = e^{-iH_{\textrm{eff}}nT}$, the evolution under the quantum walk coincides 
with the evolution under the effective Hamiltonian $U= TR_{y}(\theta)$ at integer multiple times of $T$. 
From this perspective, quantum walk provides an unique quantum simulator for the static effective Hamiltonian 
$H_{\textrm{eff}}$ through periodic drive of a quantum particle. In the following, we set $T=1$. 
While the effective Hamiltonian $H_{\textrm{eff}}$ represents a static Hamiltonian, 
there is one important difference from the truly static Hamiltonian; the energy is only defined up to $2\pi/T$,
as is clear from the definition \eqnref{effective_Hamiltonian}. This periodic structure of the eigenvalues of 
effective Hamiltonian, called quasi-energy, is the result of discrete time-translational symmetry of the system. 
Just as momentum becomes quasi-momentum with periodic structure when the system possesses only 
discrete spatial translational symmetry, energy also becomes quasi-energy with periodic structure in this case. 
This distinction between energy and quasi-energy plays an important role in understanding the topological phenomena
unique to quantum walks, or periodically driven systems, as we explain later. 

 \begin{figure}[t]
\begin{center}
\includegraphics[width = 8.5cm]{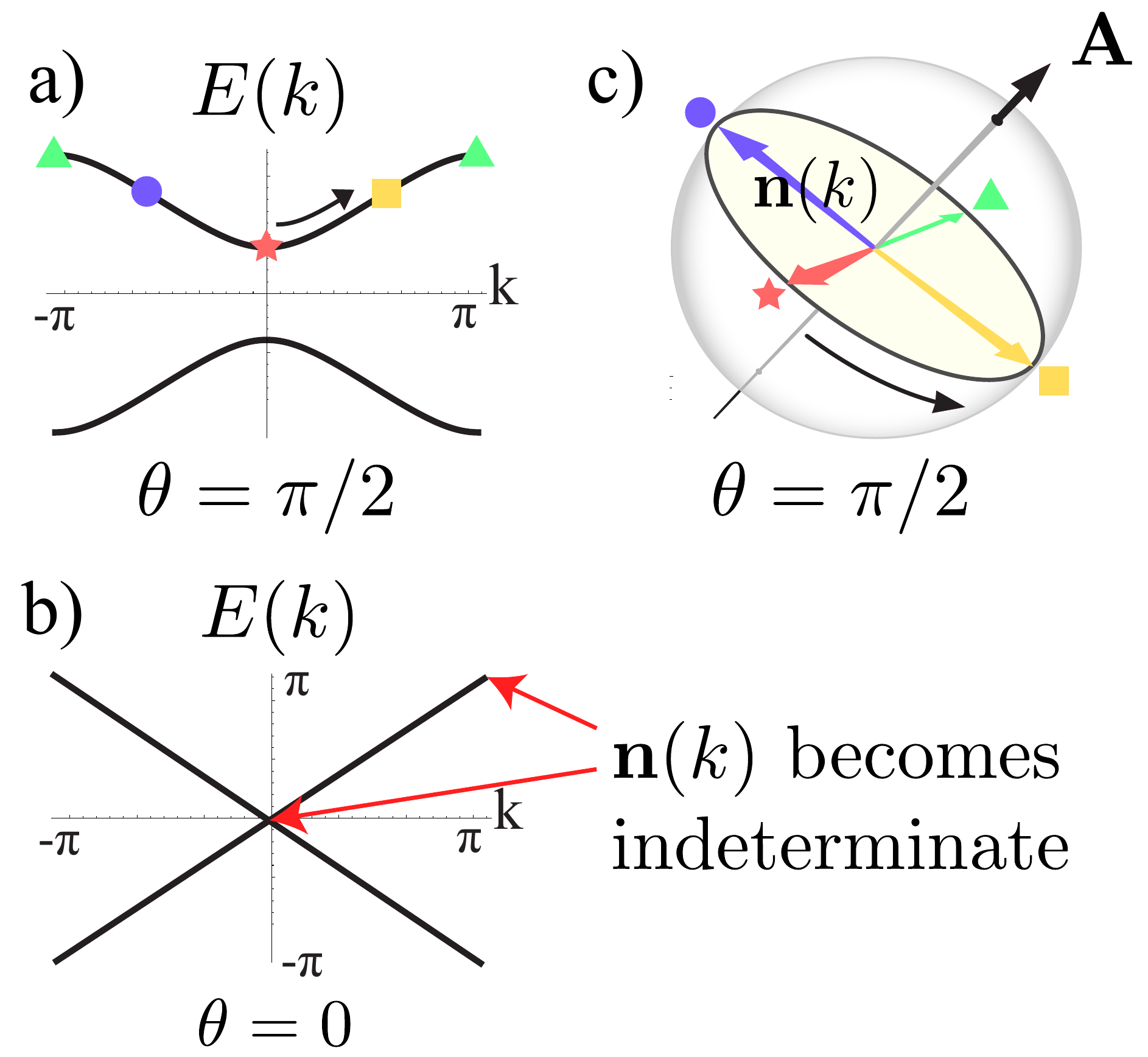}
\caption{a), b) quasi-energy spectrum of the effective Hamiltonian \eqnref{Hk} for conventional quantum walks
with rotation angle $\theta= \pi/2$ and $\theta =0$.  The spectrum consists of two bands coming from two internal 
degrees of freedom, and there is a finite gap between the two bands for general value of $\theta$ except for $\theta \neq 0, 2\pi$. For $\theta =0, 2\pi$, the spectrum closes the gap as is observed in b). Note that the 
gap is closing in this case at quasi-energy $E=0$ as well as at $E=\pi$. c) The behavior of the eigenstates $\vec{n}(k)$ 
in \eqnref{nk} represented on a Bloch sphere. For a given quasi-momentum $k$, the eigenstate is the superposition of 
spin up and down, and therefore, can be represented as a point on a Bloch sphere given by $\vec{n}(k)$. 
For a conventional quantum walk, $\vec{n}(k)$ traces a circle around the origin as $k$ goes from $-\pi$ to $\pi$. 
Note that $\vec{n}(k)$ is perpendicular to a vector $\vec{A} =( \cos(\theta/2), 0, \sin(\theta/2) )$ for all $k$ 
in our quantum walk. For gapless quantum walk with $\theta=0, 2\pi$, $\vec{n}(k)$ becomes ill-defined 
at those $k$ corresponding to the gap closing point. }
\label{fig:spectrum}
\end{center}
\end{figure}
For the quantum walk defined above, the evolution operator of one step $U= TR_{y}(\theta)$ 
possesses spatial translational symmetry, and thus the evolution operator becomes diagonal in quasi-momentum space. 
In particular, spin-dependent translation $T$ can be written as
\begin{eqnarray} 
T &=&  \sum_{j} \ket{j+1} \bra{j} \otimes \ket{\uparrow} \bra{\uparrow} + 
 \ket{j-1} \bra{j} \otimes \ket{\downarrow} \bra{\downarrow} \\
 &=&  \int^{\pi}_{-\pi} dk  \ \  e^{i k \sigma_{z}} \otimes  \ket{k} \bra{k} 
 \end{eqnarray} 
In this expression, we see that the spin-dependent translation mixes the orbital degrees of freedom represented by 
 quasi-momentum $k$ and spin encoded in $\sigma_{z}$. The presence of such spin-orbit coupling is a key to 
 realizing topological phases. We note that the continuous quantum walk does not have such spin-orbit coupling
and thus is distinct from discrete quantum walks in the topological properties\cite{Kempe2003}
 
Now in quasi-momentum space, the effective Hamiltonian for the conventional quantum walk takes the form
 \begin{equation}
\label{Hk}
H_{\textrm{eff}} = \int_{-\pi}^{\pi} dk \left[
E (k) \,\vec{n}(k) \cdot  {\bf \sigma}\right]\otimes\ket{k}\bra{k},
\end{equation}
where $\vec{ \sigma} = (\sigma_x, \sigma_y, \sigma_z)$ is the vector of Pauli matrices and the unit vector
$\vec{n}(k) = (n_x, n_y, n_z)$ defines the quantization axis for the spinor eigenstates at each momentum $k$.
For $\theta \neq 0$ or $2\pi$, explicit expressions for $E(k)$ is given by 
\begin{equation}
\label{Ek} \cos E(k) = \cos(\theta/2)\,\cos k
\end{equation}
Typical band structure of quasi-energies $E(k)$ is plotted in \fref{fig:spectrum}a) for $\theta=\pi/2$. 
There are two bands because the system has two internal degrees of freedom, and for generic values of 
$\theta$, the two bands are separated by a band gap. 

On the other hand, at $\theta=0$, $R_{y}(\theta) = I$, and therefore 
the effective Hamiltonian is $H_{\textrm{eff}} = k \sigma_{z}$. 
Thus the quasi-energy bands close the gap 
at quasi-momentum $k=0$ at quasi-energy $E=0$. 
Moreover, due to the periodicity of energy, the spectrum also closes the gap
at quasi-momentum $k=\pi$ at quasi-energy $E=\pi$.
The spectrum for $\theta=0$ is illustrated in \fref{fig:spectrum}b). 
Similar situation occurs at $\theta =2\pi$. 

Interesting structure appears in $\vec{n}(k)$ of \eqnref{Hk}.  For  $\theta \neq 0, 2 \pi$, 
$\vec{n}(k)$ is given by 
\begin{equation}
\label{nk}\vec{n}(k) =
 \frac{(\sin(\theta/2) \sin k,\  \sin(\theta/2) \cos k,\ -\cos(\theta/2) \sin k )}{\sin E(k)}. 
\end{equation}
Note that the eigenstates of the effective Hamiltonian $H_{\textrm{eff}}(\theta)$ are the superposition of 
spin up and spin down, and therefore can be represented as a point on a Bloch sphere. 
The unit vector $\vec{n}(k)$ is nothing but the unit vector that determines the direction of this point. 
The behavior of $\vec{n}(k)$ on a Bloch sphere 
 for $\theta =\pi/2$ as $k$ goes from $-\pi$ to $\pi$ is plotted in \fref{fig:spectrum}c). 
We see that $\vec{n}(k)$ "winds" around the equator of the Bloch sphere. This peculiar feature 
is in fact the origin of topological nature of quantum walks, as we see in \secref{section:topology}. 

Note that at $\theta =0, 2\pi$, the states at quasi-momentum $k=0$ and $\pi$ 
become degenerate and thus any superposition of spin up and down becomes the 
eigenstates of the Hamiltonian at these quasi-momentum. 
Thus, eigenvector $\vec{n}(k)$ becomes indeterminate at these points. 

\subsection{Hidden symmetry of quantum walks} \label{sec:symmetry}
The effective Hamiltonian of the quantum walk 
$H_{\textrm{eff}}$ as given in \eqnref{Hk} has an interesting symmetry. 
First, we note that $\vec{n}(k)$ is perpendicular to the vector 
$\vec{A} =( \cos(\theta/2), 0, \sin(\theta/2) )$ as is easy to check from \eqnref{nk}. 
The symmetry of the system is then given by the rotation around the axis $\vec{A}$ by angle $\pi$
which takes $H_{\textrm{eff}}$ to $-H_{\textrm{eff}}$, or 
\begin{equation}
\Gamma^{-1} H_{\textrm{eff}} \Gamma = -  H_{\textrm{eff}}  \quad \quad  \Gamma = e^{-i\pi \vec{A} \cdot \sigma/2}
\label{chiralsymmetry}
\end{equation}
Indeed, as is clear from the picture of  \fref{fig:spectrum}, such rotation takes $\vec{n}(k)$ to $- \vec{n}(k)$ for each $k$, and
thus takes $H_{\textrm{eff}} = \int dk E(k) \vec{n}(k) \cdot \sigma \otimes \ket{k}\bra{k} $ to minus itself. 
This symmetry given by an unitary operator is called sublattice or chiral symmetry\cite{footnote1}. 

One interesting feature of quantum walks which results from the sublattice (chiral) symmetry is 
the symmetric spectrum; states with energy $E$ and $-E$ always appear in pairs. 
This is easy to demonstrate. Given a state $\ket{\psi}$ with eigenvalue $E$, one can check that 
the state $\Gamma \ket{\psi}$ is an eigenstate of the Hamiltonian with energy $-E$;
\begin{eqnarray*}
&& H_{\textrm{eff}} \ket{\psi} = E \ket{\psi}  \\
& \rightarrow &  H_{\textrm{eff}} \left( \Gamma \ket{\psi} \right)  = -E \left( \Gamma \ket{\psi} \right)
\end{eqnarray*}
There are exceptions to this pairing of states. When $E$ satisfies $E= -E$, the states 
$\ket{\psi} $ and $\Gamma \ket{\psi}$ can represent the same state.  
Due to the periodicity of quasi-energy, quantum walk has
two special energies that satisfies the condition $E=-E$, given by $E=0$ and $E=\pi$. 
We will later see that this property of $E=0$ and $E=\pi$ leads to the topological protection of 
$E=0$ and $E=\pi$ states. 
The existence of a single $E=0$ state is known in the non-driven systems with sublattice (chiral) symmetry
\cite{Ryu}, but $E=\pi$ energy state is the novel feature of periodically driven systems.

\subsection{Asymptotic distribution; ballistic propagation} \label{sec:asymptotic}
In this section, we provide an intuition behind the propagation of a particle under quantum walks
by considering their asymptotic distributions. 
Quantum walk was originally conceived as quantum analogue of random walk\cite{Aharonov1993}. 
In fact, it is easy to check that if one carries out a measurement of the particle after each step, quantum walk reduces to 
a biased random walk. Thus, by varying the amount of decoherence in the system, one can smoothly 
go from a quantum walk to classical a random walk\cite{Broome2010}. 
However, the behavior of a random walk is quite different from that of a quantum walk in the absence of decoherence. 
We remind the reader that a (non-biased) random walk asymptotically approaches the Gaussian distribution 
with the peak centered around the origin where 
the mean squared travel distance is given by $\braket{x^2} = N a^2$ where $a$ is the step size of one step. 
Thus, the particle propagates in a diffusive fashion under a classical random walk. 

On the other hand, a particle under quantum walks propagates in a ballistic fashion\cite{Konno, Grimmett}. 
This fact almost trivially results from the understanding of quantum walks as simulations of static effective Hamiltonian 
of non-interacting particle as given by \eqnref{effective_Hamiltonian} and \eqnref{Hk}. 
If one prepares a particle in a state such that its quasi-momentum is narrowly concentrated around $k$, 
then it is intuitively clear that the particle ballistically propagates
with the group velocity given by $v_{k} = \frac{\partial E(k)}{\partial k}$ where $E(k)$ 
is the quasi-energy given in \eqnref{Ek}.

This intuition can be made rigorous by deriving the asymptotic distribution of quantum walks. 
The argument above shows that the distribution of the variable $X = x/N$ in the asymptotic limit 
converges to a finite form. Here we consider the quantum walk of a particle initially localized at the origin $x=0$ 
with spin state given by $\ket{s}$. 
In Appendix \ref{appendix:asymptotic}, we show that the distribution function of $X$ takes the following form 
\begin{eqnarray}
P(X) &=& \int^{\pi}_{-\pi} \frac{dk}{2\pi} \frac{1}{2} \left( 1+ \braket{\vec{n}(k) \cdot \sigma} \right) \delta(v_{k} - X) \nonumber \\ 
&&+ \frac{1}{2} \left( 1- \braket{\vec{n}(k) \cdot \sigma} \right) \delta(v_{k} + X)  \label{asymptotic_sol1}
\label{distribution} 
\end{eqnarray}
where $\braket{\vec{n}(k) \cdot \sigma}= \bra{s} \vec{n}(k) \cdot \sigma \ket{s}$, 
and $v_{k} = \frac{\partial E(k)}{\partial k}$. 
This result is quite intuitive; each momentum state $k$ propagates with velocity $\pm v_{k}$ where 
$+$ sign is for spin parallel to $\vec{n}(k)$ and $-$ sign is for spin anti-parallel to $\vec{n}(k)$. 
Because the initial state is localized at a single site, it is the superposition of all quasi-momentum state $k$, 
and thus, the asymptotic distribution is given by the sum of the contributions for each $k$. 

\begin{figure}[t]
\begin{center}
\includegraphics[width = 8.5cm]{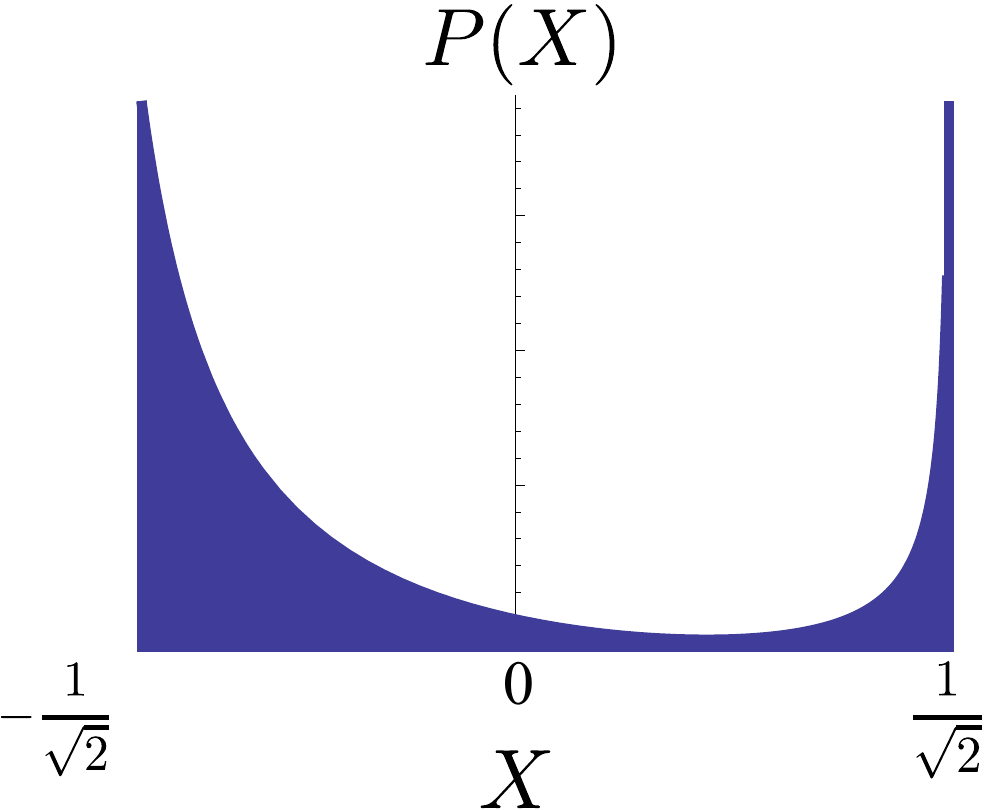}
\caption{The asymptotic distribution of a conventional quantum walk with $\theta =\pi/2$, whose analytical 
solution is obtained in \eqnref{asymptotic_sol1} and \eqnref{asymptotic_sol2}. The result is expressed 
for rescaled coordinate $X = x/N$ where $N$ is the total number of quantum walk steps. $X$ takes a finite value
for $N \rightarrow \infty$ limit because a particle propagates in a ballistic fashion. }
\label{fig:asymptotic_dist}
\end{center}
\end{figure}
The form of asymptotic distribution written above immediately leads to various results known in quantum walks
\cite{Konno, Grimmett}.
For example, symmetry of the asymptotic distribution $x \leftrightarrow -x$ exists whenever 
$\braket{\vec{n}(k) \cdot \sigma}$ is an even function of $k$, which is the case when the initial spin state is pointing
in $y$ direction. Otherwise, the distribution is generically not symmetric around the origin. 
Numerical evaluation of asymptotic distribution is always possible, and in certain cases, the analytic result 
can be expressed in a compact form. For example, for the asymptotic distribution of $\theta =\pi/2$ quantum walk
with initial spin state pointing in $z$ direction, the asymptotic distribution is given by 
\begin{equation} \label{asymptotic_sol2}
P(X) = \frac{1}{\pi} \frac{1}{1+X} \frac{1}{\sqrt{1-X^2}}  \quad  -\frac{1}{\sqrt{2}} \leq X \leq \frac{1}{\sqrt{2}}
\end{equation} 
The distribution is plotted in \fref{fig:asymptotic_dist}.
The derivation given here and Appendix \ref{appendix:asymptotic}
can be easily extended to more general quantum walks with different unitary operations 
or even to higher dimensions. 

\subsection{Experimental realizations of quantum walks} \label{sec:experiment}
One dimensional quantum walks described in previous sections 
have been realized in experiments. 
Since quantum walk is a general concept applicable to many different physical systems, 
there are realizations with cold atoms, ions, and photons\cite{Zhringer2010, Karsi2009, Schreiber2010, Broome2010}. Such realizations in different physical settings allow 
different controls of the systems, such as the ability
to choose the rotation operations, to introduce known amount of dephasing\cite{Broome2010}, or to create 
spatial boundary between regions with different rotation operations\cite{Kitagawa2011}. 
Thus, study of quantum walks in experimental settings is versatile and is not usually restricted by the technology of a specific field. 

The realization of quantum walks takes a widely different forms for different systems. 
The realization with cold atom\cite{Karsi2009} is probably the simplest and most straightforward. 
In this experiment, Karsi {\it et al} realized the quantum walk,
using the cesium atoms (Cs). 
Here, two hyperfine states of cesium (Cs) atoms are used as spin degrees of 
freedom. The spin rotation is implemented through the application of resonant microwave radiation
between these two hyperfine states. The spin-dependent translation, on the other hand, is implemented by the 
adiabatic translation of spin dependent optical lattices. This experiment implemented quantum walks up to $10$
steps, and observed the distributions of the particle. The experiment shows a good agreement, but 
quantum walks were dephased after $10$ or so steps. 

On the other hand, in the case of photonic realization by Broome {\it et al}[\onlinecite{Broome2010}], 
the vertical and horizontal polarization of a photon is used as spin degrees of freedom. 
The rotation operation is implemented through half wave plates, where the polarization of 
a photon is rotated as a photon goes through the plate. The polarization-dependent translation is implemented by 
birefringent calcite beam displacer. The optical axis of the calcite prism was cut in such way
to displace horizontally polarized light to the perpendicular direction to the propagation direction 
and transmit the vertical polarized light without displacement. 
Now these optical components are put in series in, say, $z$ direction,
so that the photon goes through them one by one as it propagates in $z$ direction. 
Thus, $z$ direction plays the role of time direction, and the photon is evolved according to the 
quantum walk as it propagates through these optical components. This experiment implemented $6$ steps.
In the subsequent experiments, this experiment was extended to create the boundary between regions with 
different rotation angles, which was used to investigate the topological nature of quantum walks\cite{Kitagawa2011}. 
Yet another implementations of quantum walks with photonic architecture is demonstrated 
in [\onlinecite{Schreiber2010}], using a fiber network loop. 

Lastly, we describe the quantum walk implemented with ions. 
In [\onlinecite{Zhringer2010}], Zhringer {\it et al} realized quantum walks in a phase space with ion $40^Ca^{+}$
They used the internal states 
$\ket{S_{1/2}, m=1/2}$ and $\ket{D_{5/2}, m=3/2}$ as spin degrees of freedom.
For spatial degrees of freedom, they used the excitation of the ions in the harmonic traps,
where the superposition of raising and lowering operators are identified as 
the coordinate operator $\hat{x} = a^{\dagger} + a$ and momentum operator 
$\hat{p} = \frac{a^{\dagger} + a}{2}$. In this space, they implemented the spin dependent translation 
by applying a bichromatic light that is resonant with both the blue and red axial sideband. 
This shows the interesting fact that quantum walks can be encoded in abstract space.
The experiment demonstrates that the quantum walk in this 
space can maintain the coherence up to even $23$ steps. 

While they are not an experimental demonstrations, there are two interesting possible 
realizations of quantum walks proposed in natural systems. 

Oka {\it et al} proposes in [\onlinecite{Oka2005}] that the evolution of electrons 
on a ring under the application of DC electric field can be understood as quantum walks in
energy level space. 
When the electric field is understood as the time-dependent vector potential, 
the problem represents a time-dependent problem, where the 
Landau-Zener transitions of electrons among different levels can be mapped to 
quantum walks. In this work, they proposed the existence of a localized state near the ground 
state of the system, which is a manifestation of the topological 
nature of quantum walks explained in this article\cite{Kitagawa2010} (see \secref{sec:reflecting}).

Another proposal by Rudner and Levitov\cite{Rudner2009} 
concerns an extension of quantum walk to include the decaying sites 
at every other lattice sites, which can arise in the problem of coupled electron 
and nuclear spins in quantum dots in the presence of competing spin-orbit and hyperfine 
interactions. This quantum walk, intriguingly, displays topological phenomena
as well, where the mean walking distance of a particle before it decays at one of the sites 
is quantized to be an integer. 

\section{Brief introduction to topological phases} \label{sec:topological_phases}
Quantum walks described in \secref{sec:1DQW} display rich topological phenomena. 
In this section, we review the ideas of topological phases, and provide the background for 
understanding the topological phenomena in quantum walks. 

The relation between quantum phases of matter and topology was first discovered through the study of 
integer quantum Hall effect, which revealed the quantization of Hall conductance for two dimensional electron gas
in the presence of a strong magnetic field\cite{Klitzing1980, MacDonald1994}. The quantization of Hall conductance is very precise, and 
moreover is robust against perturbations such as the impurity of the materials. The fundamental origin for such
robustness is the topological nature of ground state wavefunction. 
Ground state wavefunction of integer quantum Hall
systems is associated with a topological number. A topological number is a global property of the shape of wavefunction,
like "winding," and thus cannot be changed by a continuous change. Because Hall conductance is directly given by this 
topological number\cite{Thouless1982}, its value does not change under the small change of Hamiltonian or its ground state wavefunction. 

In addition to the quantized Hall conductance, yet another direct physical 
consequence of non-trivial topology of wavefunction is the existence of unidirectional edge states at the boundary of the sample\cite{Halperin1982}. 
The two phenomena, the quantized Hall conductance and unidirectional edge states, are closely related 
where the current for the quantized Hall conductance is carried by the edge states. From this point of view,
the robustness of the Hall conductance against impurities results from the robustness of unidirectional 
edge propagation against impurity scatterings. As we explain in this article, while quantized Hall conductance is 
a unique phenomenon to the topological class of integer quantum Hall systems, the existence of robust edge states 
is generic feature of any topological class. 

Because topology is a general property of ground state wavefunction, such idea is extendable to other systems 
in other dimensions\cite{Hasan2010, Qi2011}. 
In one dimension, Su Schrieffer and Heeger gave a simple model of conducting polyacetylene
and found the existence of topological solitons at edges\cite{Su1979}.
This so-called Su-Schrieffer-Heeger model of polyacetylene 
is an example of one dimensional topological phase with sublattice (or chiral) symmetry. Independently, physics in the 
same topological class was also studied in the context of high energy physics by Jackiw and Rebbi\cite{Jackiw1979, Jackiw1981}. 
In recent years, band insulators with time-reversal symmetry are predicted to possess topological properties 
called quantum spin Hall effect\cite{Kane2005, Bernevig2006}, and these so-called topological insulators were soon realized in experiments with HgTe\cite{Konig2007}. 
One important conceptual advance of these topological phases compared to integer quantum Hall phase is that 
these are new topological phases appearing in the presence of symmetries such as sublattice and time-reversal symmetry,
whereas the integer quantum Hall effect is the phenomena that appear in the absence of these symmetries. 
The idea has been further extended to three dimensional systems in the presence of time-reversal symmetry\cite{Fu2007,Chen2009, Xia2009}.

Motivated by these findings, a several groups independently classified the non-interacting systems 
in the presence of particle-hole, time-reversal, or sublattice (chiral) symmetry, giving the "periodic table"
of topological phases\cite{Schnyder2008, Qi2008, Kitaev2009}. While the possible existence
of topological classes are known for the symmetry class within these categories, 
their realizations are not easy in condensed mater materials and consequently, 
some table entry have not yet found physical realizations. 
Moreover, even when such topological phases are proposed to be realized in condensed matter materials, 
it is usually hard to directly image the wavefunction of, say, topologically protected bound states with current technology. 
Due to the outstanding controllability, artificial systems are promising alternative candidates for studying these novel phases,
and there is a number of theoretical and experimental studies of topological phenomena using cold atoms and photons
\cite{Kitagawa2011, Wang2009, Kitagawa2010b,Jiang2011, sorensen, zhu, jaksch, lewenstein}. 
Among them, quantum walks provide unique platform where any topological phase classified in one and two dimensions
is realizable with simple modifications of their protocols, as we will explain in this paper\cite{Kitagawa2010}. 
In fact, one dimensional topological class predicted to arise 
in Su-Schrieffer-Heeger model and Jackiw-Rebbi model 
has been already realized in the photonic architecture\cite{Kitagawa2011}. 
The key ingredient in the versatility of quantum walks is the controllability of the protocols. Because protocols 
are something experimentalists choose to implement, 
it is possible to design the protocols in such a way to preserve or break
certain kind of symmetries. From the following section, we study how such topological structure appears in 
discrete quantum walks.

\section{Topological phenomena in quantum walks} \label{section:topology}

\subsection{Topological characterization of quantum walks} \label{sec:winding}
In \secref{sec:intro}, we have seen that the eigenvectors of quantum walks 
$\vec{n}(k)$ illustrated in  \fref{fig:spectrum} have non-zero winding 
as $k$ goes from $-\pi$ to $\pi$. Such winding gives a topological characterization of 
quantum walks in the presence of sublattice (chiral) symmetry.

Aas we noted in \secref{sec:intro}, the effective Hamiltonian of the quantum walk possesses 
the sublattice (chiral) symmetry, which constrains the eigenvector of Hamiltonian for each quasi-momentum 
 $\vec{n}(k)$ to lie on the plane perpendicular to the vector $\vec{A}$. 
Under this sublattice (chiral) symmetry,  $\vec{n}(k)$ represents a map from the first Brillouin zone,
 which is a circle, to the equator of Bloch sphere, which is also a circle. Then quantum walk described by \eqnref{Hk} 
 possesses non-trivial winding of this map, where $\vec{n}(k)$ goes around once the equator of Bloch sphere as 
 $k$ goes from $-\pi$ to $\pi$.  
 In the presence of the symmetry, the winding observed in  \fref{fig:spectrum} c) is robust against small perturbations; one cannot change the winding number by small change of the Hamiltonian $H_{\textrm{eff}}$ which preserves the sublattice (chiral) symmetry. 
One can intuitively check this robustness by trying to change the winding through the continuous deformations
of $\vec{n}(k)$. 
 We call this winding number as topological number due to their robustness against perturbations, and 
 the topological number associated with $H_{\textrm{eff}}$ is $1$ whenever $\theta \neq 0, 2\pi$\cite{footnote2} for our
 quantum walks. 
More generally, the winding of  $\vec{n}(k)$ around the equator can take any integer value, and 
 different topological phases in this topological class 
are associated with different integers (winding numbers), and thus the topological classification is given by ${\bf Z}$
(a set of integers).

It is important to note that this winding number is not topological in the absence of any symmetry constraint,
 in the sense that the winding number can be made zero by continuous change of $\vec{n}(k)$.
 For example, one can shrink the loop of $\vec{n}(k)$ into a point on a Bloch sphere. 

 \begin{figure*}[t]
\begin{center}
\includegraphics[width = 13cm]{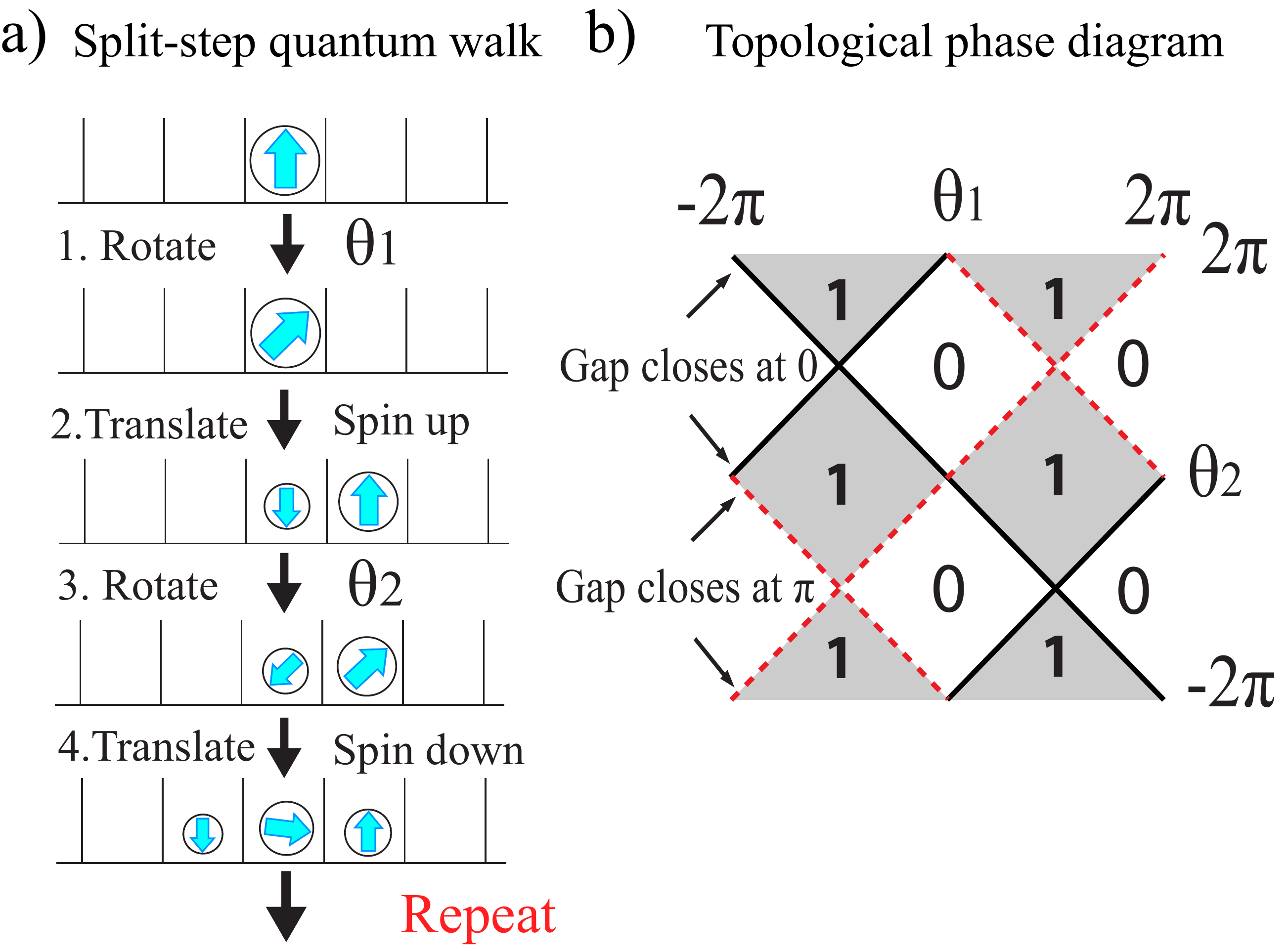}
\caption{a) Protocol for a split-step quantum walk. Split-step quantum walk is defined for spin $1/2$ particle
in one dimensional lattice, and consists of four operations; 1. spin rotation around $y$ axis by angle $\theta_{1}$, given by $R_{y}(\theta_{1})$; 2. Translation of spin up to the right by one lattice, $T_{\uparrow}$; 3. spin rotation around $y$ axis by angle $\theta_{2}$, given by $R_{y}(\theta_{2})$; 4. Translation of spin down to the left by one lattice, $T_{\downarrow}$. 
The evolution operator of one step is given by $U = T_{\downarrow} R_{y}(\theta_{2}) T_{\uparrow} R_{y}(\theta_{1})$. 
b) The topological phase diagram of split-step quantum walk for various rotation angles $\theta_{1}$ and $\theta_{2}$. 
The phase is characterized by the winding number $W$, and split-step quantum walks realize either $W=0$ or $W=1$. 
Since winding number is a topological number, it can change its value only when the band gap closes, and these
gapless phases are denoted by solid black line (band gap closes at quasi-energy $E=0$) and by red dotted line
(band gap closes at quasi-energy $E=\pi$). }
\label{fig:split_step}
\end{center}
\end{figure*}

More generally, the concept of topological numbers (or topological invariants) are defined for a collection of 
Hamiltonian that represent band insulators with certain symmetries. 
 A topological invariant is assigned to each band, 
 and its value cannot change under the continuous deformations of Hamiltonian which preserve 
the symmetry. There is one exception to this statement; when two bands mix with each other, the topological 
invariants can be changed in these two bands. We can flip this argument and say that 
topological numbers can change their values only if two bands close their band gaps in the process. 
We argue in \secref{sec:bound_state} that this property of topological numbers results in the creation of 
topologically protected bound states in the spatial boundary between regions that belong to two distinct 
topological phases. 

In the conventional quantum walk, the winding number associated with the effective Hamiltonian is always 
$1$, and no other topological phase exist in this family of quantum walks. 
In the next section, \secref{sec:splitstep}, 
we give yet another family of quantum walks that display two distinct topological phases
with winding number $0$ and $1$ in the presence of sublattice (chiral) symmetry. 
We illustrate how topological numbers can change as Hamiltonian is changed in this example.

\subsection{Split step quantum walks} \label{sec:splitstep}
In this section, we extend the conventional quantum walk by modifying the protocols 
and define so-called split-step quantum walks. This example illustrates how one can 
engineer topological phases in quantum walks through the active design of the protocols.

Split-step quantum walks is a simple extension of conventional quantum walks which have one additional rotation and translation process(see \fref{fig:split_step} a)). The complete protocol is as follows; 
\begin{enumerate}
\item Rotation of the spin around $y$ axis by
angle $\theta_{1}$, corresponding to the operation $R_{y}(\theta_{1})= e^{-i\theta_{1} \sigma_{y}/2}$. 
\item Translation of spin up particle to the right, given by 
$T_{\uparrow} = \sum_{j} \ket{j+1} \bra{j} \otimes \ket{\uparrow} \bra{\uparrow} 
+ {\bf 1} \otimes \ket{\downarrow} \bra{\downarrow}$.  Spin down particle stays in the original position. 
\item Second rotation of the spin around $y$ axis by
angle $\theta_{2}$, corresponding to the operation $R_{y}(\theta_{2})= e^{-i\theta_{1} \sigma_{y}/2}$. 
\item Translation of spin down particle to the left, given by 
$T_{\downarrow} = \sum_{j} \ket{j-1} \bra{j} \otimes \ket{\downarrow} \bra{\downarrow} 
 + {\bf 1} \otimes \ket{\uparrow} \bra{\uparrow}$.  Spin up particle stays in the original position. 
\end{enumerate}
Thus, the evolution operator of one step is given by 
$U(\theta_{1}, \theta_{2}) = T_{\downarrow} R_{y}(\theta_{2}) T_{\uparrow}R_{y}(\theta_{1})$.
This split-step quantum walk is reduced to the conventional quantum walk defined in \secref{sec:intro}
with $\theta_{2}=0$. 
As before, we can find the effective Hamiltonian through $U \equiv e^{-iH_{\textrm{eff}}}$. 
The effective Hamiltonian again takes the form \eqnref{Hk} where the quasi-energy is 
\begin{equation}
\cos E(k) = \cos(\theta_{2}/2) \cos(\theta_{1}/2) \cos k - \sin(\theta_{1}/2)\sin(\theta_{2}/2),  
\end{equation}
and the eigenvector $\vec{n}(k)$ is 
\begin{eqnarray} \label{splitstepn}
n_{x}(k) &=&  \frac{\cos(\theta_{2}/2) \sin(\theta_{1}/2) \sin k}{\sin E(k)} \nonumber \\
n_{y}(k) &=&  \frac{\sin(\theta_{2}/2)\cos(\theta_{1}/2)+ \cos(\theta_{2}/2)\sin(\theta_{1}/2) \cos k}{\sin E(k)} \nonumber \\
n_{z}(k) &=&  \frac{-\cos(\theta_{2}/2) \cos(\theta_{1}/2) \sin k }{\sin E(k)}.  \nonumber
\end{eqnarray}
It is straightforward to check that $\vec{A}(\theta_{1})= (\cos(\theta_{1}/2),\, 0,\, \sin(\theta_{1}/2))$
is perpendicular to $\vec{n}(k)$ for all $k$.
Therefore, the system possesses sublattice (chiral) symmetry with the symmetry operator
$\Gamma(\theta_{1})= e^{-i\pi \vec{A}(\theta_{1}) \cdot {\bm \sigma}/2}$.
Notice that the this symmetry operation only depends on the first rotation angle $\theta_{1}$. 

The existence of sublattice (chiral) symmetry allows us to characterize the split-step quantum walk by
the winding number, denoted by $W$, of $\vec{n}(k)$ around the equator of Bloch sphere. 
Using the explicit expression for $\vec{n}(k)$ in Eq. (\ref{splitstepn}), we find $W = 1$ if
$|\tan(\theta_{2}/2)/\tan(\theta_{1}/2)| <1$, and $W = 0$ if $|\tan(\theta_{2}/2)/\tan(\theta_{1}/2)| >1$.
Thus the split-step quantum walk can realize different winding number for different rotation angles
$\theta_{1}$ and $\theta_{2}$. 
We plot the phase diagram of split-step quantum walk in \fref{fig:split_step} b).


For a given $\theta_{1}$, a set of Hamiltonian for varying values of second rotation $\theta_{2}$
has the same sublattice (chiral) symmetry, and thus 
$\{ H_{\theta_{1}}(\theta_{2}) \}$ defines quantum walks in the same topological class. Because the dependence
of $H_{\theta_{1}}(\theta_{2})$ on $\theta_{2}$ is continuous, topological nature of the winding number implies that 
the winding number is the same for wide range of the rotation angle $\theta_{2}$ as 
phase diagram  \fref{fig:split_step} b) shows. 

However, the winding number can change its value when the two bands close their gap.
Such gapless points are given by the points $|\tan(\theta_{2}/2)/\tan(\theta_{1}/2)| = 1$, 
or $\theta_{2} = \pm \theta_{1}, 2\pi \pm \theta_{1}$ denoted by solid and dotted lines in  \fref{fig:split_step} b). 
The mechanism behind the change of winding numbers is the following.
At the value of $\theta_{2}$ where two bands close the gap, for example, $\theta_{2}=\theta_{1}$, 
eigenvector $\vec{n}(k)$ in \eqnref{splitstepn} 
becomes ill-defined at the quasi-momentum $k$ corresponding to the degenerate points,
because any superposition of spin up and down is the eigenstate of the Hamiltonian at that points. 
At this $\theta_{2}= \theta_{1}$, the winding number cannot be defined, 
and winding numbers of bands at $H_{\theta_{1}}(\theta_{1}-\epsilon)$ and 
$H_{\theta_{1}}(\theta_{1}+\epsilon)$ do not have to be the same. 

Note that this argument can be used to find the phase diagram in \fref{fig:split_step} b) without
calculating the winding number at each point of the phase diagram. 
Since the winding number can change only when the bands close their gap,
the phase transition between two topological phases is always gapless. 
Thus, in order to draw the phase diagram, it is only necessary to identify the gap closing points
in the parameter space, and find the winding number of the region bounded by the gapless phase lines. 
In quantum walks, the periodicity of quasi-energy allows the closing of the gap at either
quasi-energy $E=0$ and $E=\pi$ as we saw in \fref{fig:spectrum} b). 
These gap closing lines at quasi-energy $E=0$ and $E=\pi$ are denoted in  \fref{fig:split_step} b) as
solid and dotted lines, respectively. 

The topological structure of split-step quantum walk described above 
has a strong asymmetry between $\theta_{1}$ and $\theta_{2}$, but this asymmetry is an artifact. 
One can shift the starting time of the quantum walk by unitary transformation and define 
an equivalent dynamics through $U' = T_{\uparrow}R_{y}(\theta_{1})T_{\downarrow} R_{y}(\theta_{2}) $.
It is straightforward to show that this quantum walk has sublattice (chiral) symmetry given by 
$\Gamma(\theta_{2})= e^{-i\pi \vec{A}(\theta_{2}) \cdot {\bm \sigma}/2}$ with 
$\vec{A}(\theta_{1})= (\cos(\theta_{1}/2),\, 0,\, \sin(\theta_{1}/2))$. In this case, 
the quantum walks with constant $\theta_{2}$ correspond to the Hamiltonians 
in the same topological classes.

\subsection{Physical manifestations of \\ topological band structure} \label{sec:bound_state}
The non-trivial winding, or topological number, of the bands in quantum walks gives rise to 
a robust bound states at the boundary between two phases with different topological numbers. 
In the following, we first give an intuition behind the existence of such robust edge states. 

 \begin{figure}[t]
\begin{center}
\includegraphics[width = 8.5cm]{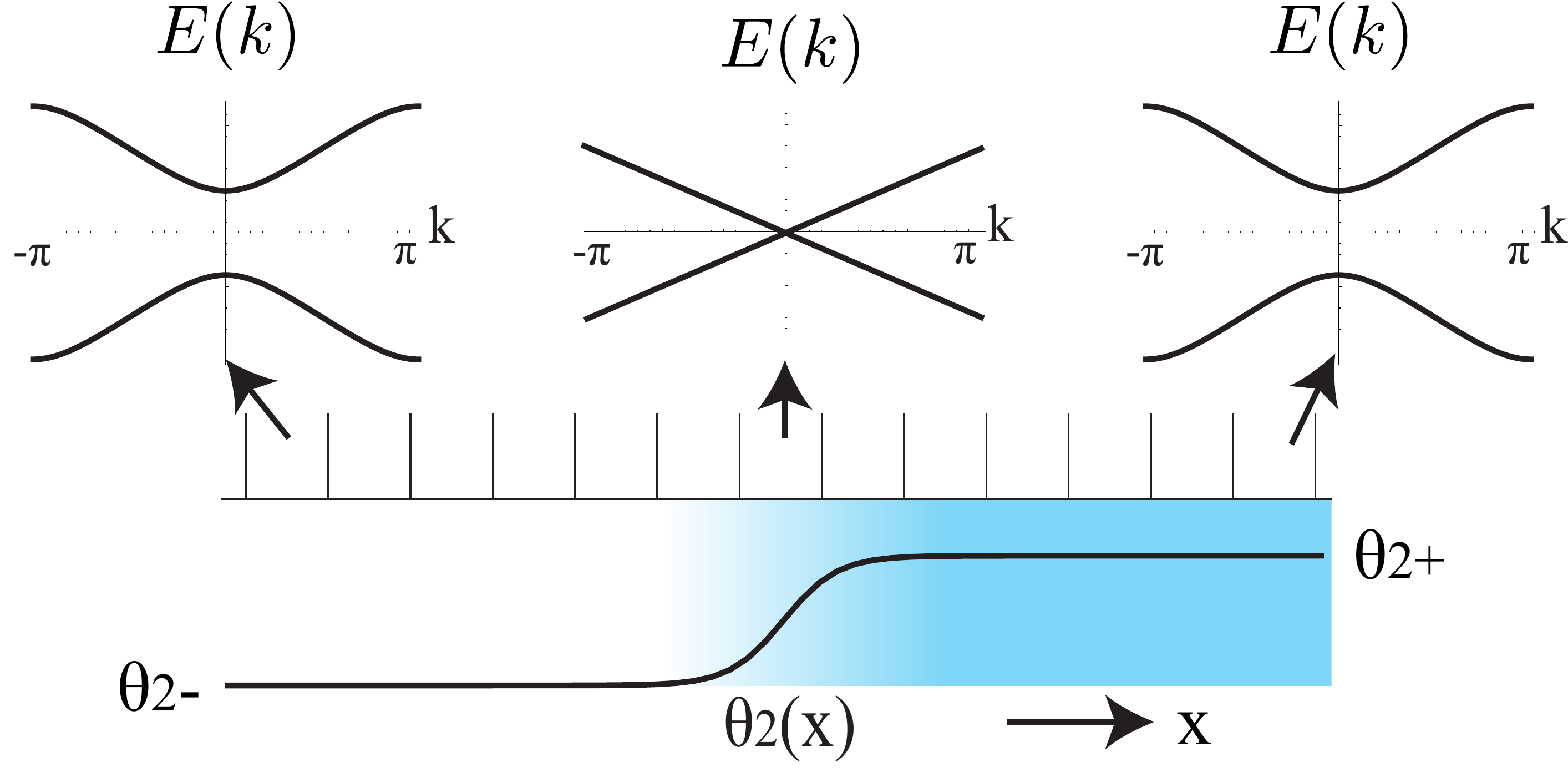}
\caption{Illustration of the existence of bound states across the boundary of regions 
that belong to distinct topological phases. Here, we consider the inhomogeneous split-step quantum walk
where the second rotation $\theta_{2}(x)$ changes in space, and the winding number associated with 
the phase $\theta_{2-} = \theta_{2}(x \rightarrow -\infty)$ is different from the winding number 
associated with the phase $\theta_{2+} = \theta_{2}(x \rightarrow \infty)$. In both limit of $x \rightarrow -\infty$
and $x \rightarrow \infty$, the bands are gapped. However, the winding number cannot change its value 
unless band gap closes, and thus it is expected that the band gap closes in the middle near the origin. 
States at $E=0$ that exist near $x=0$ must be localized since there is no state at this energy far into the left 
or into the right of the system.  }
\label{fig:boundary_scheme}
\end{center}
\end{figure}

Here we consider the split-step quantum walks with inhomogeneous rotations in space, in order
to create a boundary between quantum walks with different winding numbers. 
Here we take the first rotation $\theta_{1}$ to be 
homogeneous in space and make the second rotation $\theta_{2}$ space dependent, 
see \fref{fig:boundary_scheme}. 
The effective Hamiltonian of this inhomogeneous quantum walk, 
having the homogeneous first rotation $\theta_{1}$, possesses the sublatice (chiral) symmetry
given by $\Gamma(\theta_{1})= e^{-i\pi \vec{A}(\theta_{1}) \cdot {\bm \sigma}/2}$. 
While this statement is intuitively clear, it is instructive to explicitly show the existence of sublattice 
(chiral) symmetry, and we provide the proof in the Appendix \ref{appendix:sublatice_symmery}. 

We take the second rotation angle to approach $\theta_{2-}$ for $x \rightarrow -\infty$
and $\theta_{2+}$ for $x \rightarrow \infty$. We take the region in which the rotation angle changes 
from $\theta_{2-}$ to $\theta_{2+}$ to be finite region around $x=0$. 
Now we take $\theta_{2-},\theta_{2+}$ to be such that
winding number is $0$ for the split-step quantum walk with the rotation angle $\theta_{1} ,\theta_{2-}$ and 
winding number is $1$ for the walk with the rotation angle $\theta_{1} ,\theta_{2+}$. 
Then the region near the origin represents the phase boundary between two distinct topological phases. 

 \begin{figure*}[th]
\begin{center}
\includegraphics[width = 14cm]{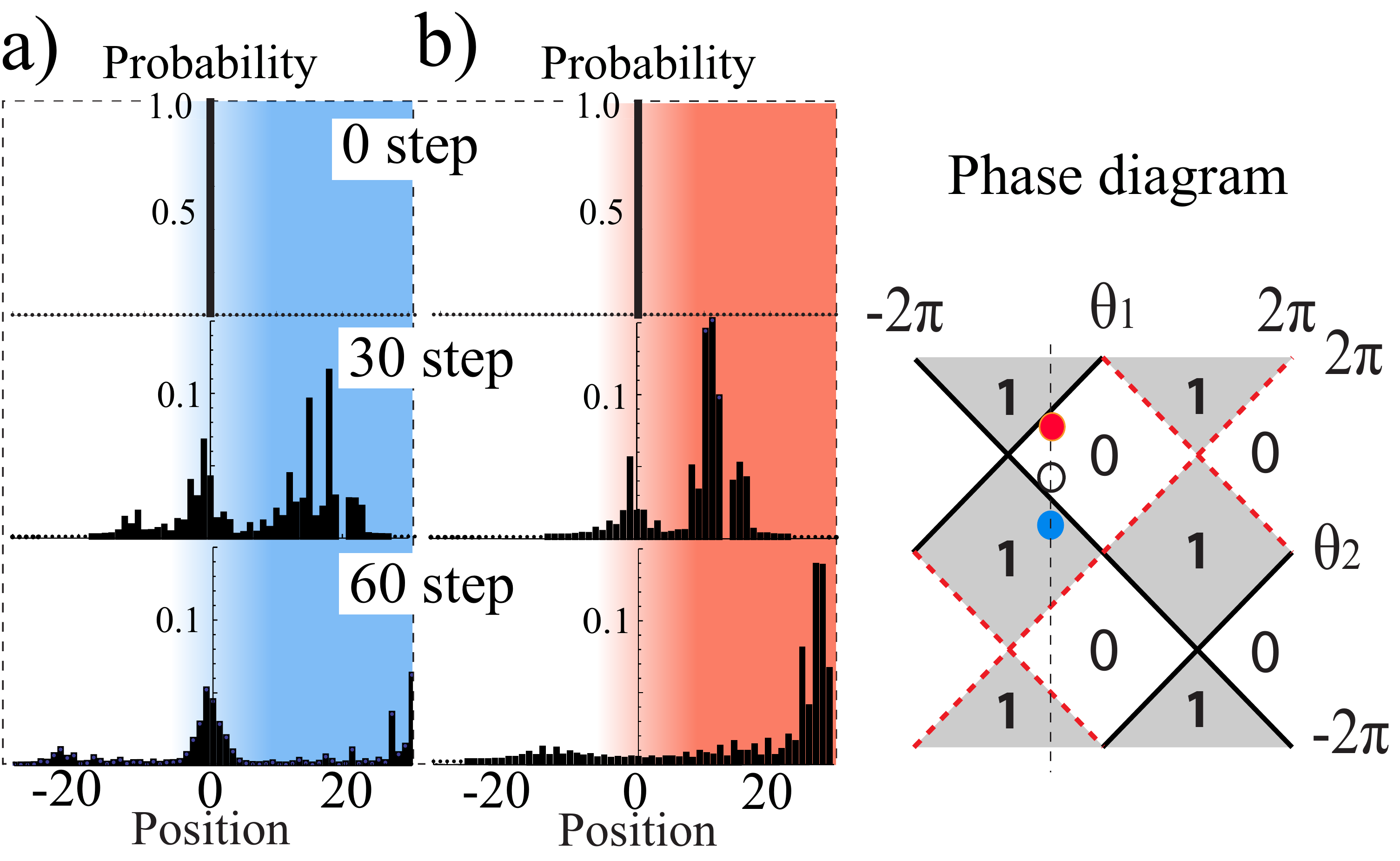}
\caption{ Evolution of the spatially inhomogeneous split-step quantum walk.
The initial spin of the particle is spin up, and particle is initialized at $x = 0$. 
a) The rotation angles of quantum walk are chosen such that the quantum walk corresponds to 
trivial topological phase with winding number $W=0$ as $x \rightarrow -\infty$ indicated as a white dot in the phase diagram 
and non-trivial topological phase with $W=1$ as $x \rightarrow \infty$ indicated as a blue dot.  
Here we took the uniform first rotation $\theta_1 = -\pi/2$ and second rotation $\theta_{2-} = 3\pi/4$ and $\theta_{2+} = \pi/4$
with $\theta_{2}(x) = \frac12 (\theta_{2-}+ \theta_{2+})+ \frac12(\theta_{2+}- \theta_{2-}) \tanh (x/3)$. 
After many steps of quantum walks, a large probability of finding a particle near the origin remains, indicating the 
presence of bound states. 
b) In this quantum walk, 
the phase of the two sides of the origin has the same winding number. 
The phase as $x \rightarrow -\infty$ is indicated by the white dot in the phase diagram and 
phase as $x \rightarrow \infty$ is indicated by the red dot. 
Here we took  $\theta_1 = -\pi/2$ and $\theta_{2-} = 3\pi/4$ as before and $\theta_{2+} = 11\pi/8$. 
In this case, the probability to find the walker near $x = 0$ after many steps decays to 0, indicating the absence of a localized state at the boundary.}
\label{fig:bound_state}
\end{center}
\end{figure*}

In \fref{fig:boundary_scheme}, we illustrate the local band structures in this inhomogeneous quantum walk. 
Strictly speaking, the system is spatially inhomogeneous, so quasi-momentum is no longer a good quantum number,
and band structures cannot be drawn. 
Yet, it is helpful to visualize band structures to understand what happens at the boundary. 
If the variation of rotation angle $\theta_{2}$ is slow, then one can visualize the band structures at point $x_{0}$ to be
the band structures of homogeneous quantum walk with rotation angles $\theta_{1}$ and $\theta_{2}(x_{0})$. 
Such description is certainly applicable in the limit $x \rightarrow -\infty$ and $x \rightarrow \infty$. 

The band structures of both $x \rightarrow -\infty$ and $x \rightarrow \infty$ represent band insulators 
where two bands are separated by a band gap. By definition, the two bands are characterized by 
winding number $W=0$ in $x \rightarrow -\infty$ and $W=1$ in $x \rightarrow \infty$.
However, since the sublattice (chiral) symmetry exists throughout the space, the winding numbers can only change
by closing the gap across the boundary near $x=0$. Therefore, the gap must close near $x=0$, as illustrated in \fref{fig:boundary_scheme}. 

This argument shows that there must be states in the gap (near $E=0$) around the origin. 
Now because there is no state near $E=0$ in the limit $x \rightarrow \infty$ and $x \rightarrow -\infty$ 
(this energy is in the gap of the bands), we can conclude these states near $E=0$ around the origin
must be confined around the origin. Thus there is generically a bound state
near $E=0$ at the boundary between two different topological phases. 

This prediction can be confirmed by running a simple simulations of inhomogeneous quantum walks.
The presence or absence of bound states can be confirmed by initializing the particle 
near the origin and running the quantum walk protocols. If there are bound states near the origin, 
there is generically an overlap between the initial state and the bound state, and
even after many steps of quantum walk, there remains a
non-zero probability to find the particle near the origin. On the other hand, 
if there is no bound state, the particle quickly propagates away from the origin due to the ballistic propagation of
quantum walks as described in \secref{sec:asymptotic}. 

In \fref{fig:bound_state}, we present the result of two inhomogeneous quantum walks.
In  \fref{fig:bound_state} a), the boundary between two topologically distinct phases is created near the origin 
with winding number $W=0$ as $x \rightarrow -\infty$ and $W=1$ as $x \rightarrow \infty$. 
Specifically, the uniform first rotation is $\theta_1 = -\pi/2$ 
and second rotation $\theta_{2-} = 3\pi/4$ and $\theta_{2+} = \pi/4$. 
Here we considered a smooth variation of the second rotation given by 
$\theta_{2}(x) = \frac{1}{2} (\theta_{2-}+ \theta_{2+})+ \frac{1}{2}(\theta_{2+}- \theta_{2-}) \tanh (x/3)$,
where the second rotation changes from $\theta_{2-}$ to $\theta_{2+}$ with the length scale of $\sim 6$ sites. 
The phases of quantum walks in the limit $x \rightarrow -\infty$ and $x \rightarrow \infty$ are indicated on the 
phase diagram as the white and blue dot, respectively. 
In the simulation,  the initial spin of the particle is spin up, and particle starts at $x = 0$. 
As we expect, a peak in the probability distribution appears even after $60$ steps of the simulation, indicating 
the existence of topological bound states. 

On the other hand, we studied the quantum walk in \fref{fig:bound_state} b), where the system is characterized by 
$W=0$ throughout the space. Here we took $\theta_1 = -\pi/2$,  $\theta_{2-} = 3\pi/4$ and $\theta_{2+} = 11\pi/8$
again with the same functional dependence on $x$, 
$\theta_{2}(x) = \frac{1}{2} (\theta_{2-}+ \theta_{2+})+ \frac{1}{2}(\theta_{2+}- \theta_{2-}) \tanh (x/3)$.
The phases of quantum walks in the limit $x \rightarrow -\infty$ and $x \rightarrow \infty$ are indicated on the 
phase diagram as the white and red dot, respectively. 
With the same initial state of spin up, we implemented the quantum walk simulation, and the resulting 
probability distribution shows a fast decay of probability near the origin. After $60$ steps, the probability near
the origin decays close to zero, indicating the absence of bound states. 

For further details of the simulation, the interactive demonstration of inhomogeneous quantum walks is available 
on the Mathematica demonstration website\cite{mathematica, footnote3}, where 
one can change the values of $\theta_{1}$ and $\theta_{2}$ and run the quantum walks.

The topological class realized by the conventional quantum walk and by the split-step quantum walk
 is the same topological class as is proposed in 
 Su-Schrieffer-Heeger model of polyacetylene and Jackiw-Rebbi model. 
 This is the topological class in one dimension with sublattice (chiral) symmetry. 
In this respect, quantum walk acts as a quantum simulator of the topological phase.
Since quantum walks are realizable with many different systems such as ions, photons and cold atoms, 
they allow the study of topological phases in a manner that is not possible in traditional condensed matter materials. 
 The proposal to study topological phases and topologically protected bound states in split-step quantum walk 
 was first proposed in [\onlinecite{Kitagawa2010}], and later realized in experiments with photonic architecture\cite{Kitagawa2011}.
 The controllability of the experimental apparatus not only allow the direct imaging of the wavefunctions of topological bound states, but also allow to confirm the robustness of the bound states with parameter changes, the signature of topological origin for the bound states.  This topological bound state in one dimensional system has not been directly 
 observed in materials such as polyacetylene, and this photonic architecture provided the first experimental
 imaging of the bound states with topological origin in one dimension.

\subsection{Quantum walks with a reflecting boundary} \label{sec:reflecting}
A special case of Hamiltonian with trivial topology is the vacuum, where the topological number associated with
the system is zero. 
Thus, we can make a phase boundary by simply terminating the quantum walk with winding number $W=1$.
In this setup, the boundary exists at the edge of the quantum walk, and it is expected that a bound
state exists at this edge according to the general argument in \secref{sec:bound_state}. 

As we briefly mentioned in \secref{sec:experiment}, 
such a setup is not unphysical, and intriguing realization of quantum walks with a reflecting edge 
has been suggested in [\onlinecite{Oka2005}] through the understanding of a particle under the electric field 
in discrete energy level structure as a quantum walk. The transition between different levels occurs as 
Landau-Zener process in this system, which corresponds to the translation operation in quantum walks. 
Thus, the ground state of the system acts as the reflecting boundary. They predicted the existence of a bound 
state near the ground state, which we can now understand as a topological bound state as a result of 
non-trivial topological number of quantum walks. 

Here we consider the conventional quantum walk described by $U = T R_{y}(\theta)$ which extends from 
$x = - \infty$ up to $x=0$. The quantum walk is terminated at $x=0$. In order to conserve the particle number,
we require the operation at the boundary to be unitary, {\it i.e.} the spin $\uparrow$ particle needs to be reflected 
at the edge $x=0$. 
Here we take the following operation at the edge $x=0$; 
\begin{enumerate}
\item Rotation of the spin around $y$ axis by
angle $\theta$, as in other sites, given by $R_{y}(\theta)= e^{-i\theta \sigma_{y}/2}$. 
\item Translation of the spin $\downarrow$ to site $x=-1$.  Spin $\uparrow$ stays at $x=0$
and its spin is flipped to spin $\downarrow$ with phase accumulation $e^{i\phi}$
\end{enumerate}
Explicitly, the operation at $x=0$ is 
\begin{eqnarray*} 
U(x=0)  &=& T_{\textrm{edge}} R_{y}(\theta) \\
T_{\textrm{edge}} &=& \ket{-1}\bra{0} \otimes  \ket{\downarrow}\bra{\downarrow} + 
e^{i\phi} \ket{0}\bra{0} \otimes  \ket{\downarrow}\bra{\uparrow} 
\end{eqnarray*}

In order to have a quantum walk in a topological class, it is crucial to have the sublattice (chiral) symmetry
of the whole system. In particular, the sublattice (chiral) symmetry needs to be present for the evolution 
operator including the edge. If we denote the total evolution of the system with an edge 
as $U_{x \leq 0}$, then we require 
\begin{equation}
\Gamma^{-1} U_{x \leq 0} \Gamma = U_{x \leq 0}^{\dagger}
\end{equation} 
where $\Gamma{=}e^{-i\pi \vec{A} \cdot \vec{\sigma}/2}$ with 
$\vec{A} =[ \cos(\theta/2), 0, \sin(\theta/2) ]$. This is a simple extension of the definition
of sublattice (chiral) symmetry in \eqnref{chiralsymmetry} to evolution operator. 
It is straightforward to check that the necessary and sufficient condition for the existence of 
sublattice (chiral) symmetry in $U_{x \leq 0}$ is $\phi =0, \pi$ for the phase accumulated at the reflecting boundary. 

According to the general argument in \secref{sec:bound_state}, bound states exist near the boundary
of $x=0$. For this simple quantum walk, it is not very difficult to obtain the analytical solution of the 
bound state. The details of the derivation is given in the Appendix \ref{appendix:bound_state}. 
We note that a straightforward extension of the derivation given in the Appendix should allow 
similar analytical solutions of bound states for the inhomogeneous split-step quantum walks. 

Here we describe the solution for the boundary condition of $\phi=0$ and rotation angle $\theta = \pi/2$.
The analytical solution shows that the bound state is at quasi-energy $E=\pi $ and the wavefunction takes the form 
\begin{eqnarray}
\label{bound_state} \ket{\psi_{b}^{E=0}(-j)} &=& \frac{1}{\mathcal{N}} (-1)^j e^{- j/\lambda} \otimes \left( \begin{array}{c}1 - \sqrt{2} \\ 1  \end{array}  \right)
\quad  0 \leq j \nonumber \\
&& \\
\lambda &=& - \frac{1}{\log(\sqrt{2}-1)} \nonumber
\end{eqnarray}
where $\mathcal{N}$ is the normalization factor. 
Since the localization length $\lambda \approx 1.1$ is small, this bound state 
is tightly localized around $x=0$. 

For the same boundary condition $\phi=0$ but different rotation angle $\theta = 5\pi/2$, 
the evolution operator $U = T R_{y}(\theta)$ is different from the one with the rotation angle $\theta = \pi/2$
by only a minus sign, {\it i.e.} $ T R_{y}(\theta = 5\pi/2) = - T R_{y}(\theta = \pi/2)$. Thus, the same wavefunction 
in \eqnref{bound_state} is the bound state for this case as well, but now the quasi-energy of the bound state
is $E=0 $ due to the extra minus sign. 

On the other hand, for the reflecting boundary condition with phase accumulation $\phi=\pi$ with $\theta = \pi/2$,
the bound state exists at quasi-energy $E= 0 $ and wavefunction is
\begin{eqnarray*}
\ket{\psi_{b}^{E=\pi} (-j)} &=& 
\frac{1}{\mathcal{N}} e^{- j/\lambda} \otimes \left( \begin{array}{c}1 - \sqrt{2} \\ 1  \end{array}  \right)
\quad  0 \leq j \\
\lambda &=& - \frac{1}{\log(\sqrt{2}-1)}
\end{eqnarray*}

Generally, the quasi-energy of bound states for quantum walks with sublattice (chiral) symmetry 
is at $E=0$ or $E=\pi$. We will see in the next section, \secref{sec:topological_invariant}, that these 
energies represent special points where sublattice (chiral) symmetry provides 
the topological protection of the states at these energies. 

For the special rotation angles of $\theta=\pi$, it is possible to obtain the bound state solution by following the quantum walk 
for a few steps. Since this calculation is elementary, the existence of a bound state can be easily understood. This rotation angle corresponds to the rotation operation
\begin{eqnarray*}
R_{y}(\theta =\pi)& =& -i \sigma_{y} \\
& =&    \left( \begin{array}{cc} 0 & -1 \\  1 & 0  \end{array}  \right)
\end{eqnarray*}
Thus, the rotation turns $\ket{\uparrow} \rightarrow \ket{\downarrow} $ 
and  $\ket{\downarrow} \rightarrow - \ket{\uparrow} $. 
Let us take the phase accumulation upon reflection to be $\phi=0$. 
If we initialize the particle at $x=0$ with spin down, the quantum walk follows the following evolution.
\begin{eqnarray*}
\ket{0} \otimes \ket{\downarrow} 
\stackrel{R}{\rightarrow } \ket{0} \otimes \left( - \ket{\uparrow} \right)
\stackrel{T}{\rightarrow } \ket{0} \otimes  \left(- \ket{\downarrow} \right)
\end{eqnarray*}
Thus in this special case, $\ket{0} \otimes \ket{\downarrow} $ is an eigenstate of quantum walk operator 
localized at the edge. Since the state gains a minus sign after one step of the quantum walk,
the quasi-energy of the bound state is $\pi$, in accordance with the result obtained in the general analytical solution.

\subsection{Topological protection of the bound states:  topological invariants} \label{sec:topological_invariant}
The bound states resulting from topology studied in \secref{sec:bound_state} and 
 \secref{sec:reflecting}  are protected in a sense that 
they are robust against small changes in the quantum walk protocols. The logic is the following; since the 
bound states are the result of topological winding numbers, and topological winding numbers cannot 
change their values unless the bands close their gaps, the bound states cannot disappear 
for a small change of rotation angles unless they are changed by a large amount 
such that the gap of the corresponding effective Hamiltonian closes. 

 \begin{figure}[t]
\begin{center}
\includegraphics[width = 8.5cm]{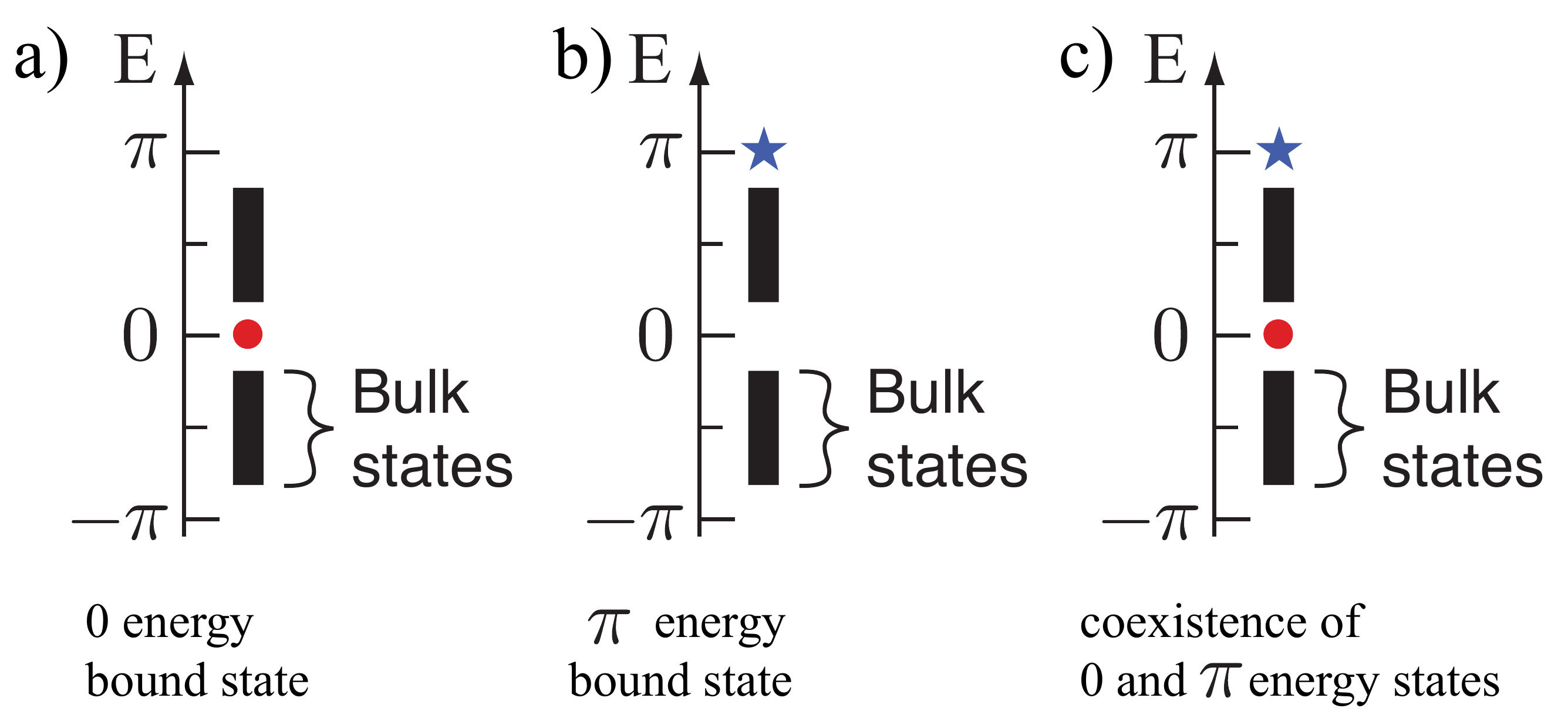}
\caption{General spectrum structure for inhomogeneous quantum walks with bound states. 
a), b) in \secref{sec:reflecting}, we analytically showed that there is a $0$ or $\pi$ energy bound state
at the boundary of a conventional quantum walk. In a similar fashion, these bound states also appear in the inhomogeneous split-step quantum walks. These bound states energy are well separated from 
the extended, bulk states. Sublattice (chiral) symmetry requires that states with energy $E$ and $-E$ appear 
in pairs, and thus, a single $0$ and $\pi$ energy bound state cannot disappear unless bulk energy gap closes
at $0$ or $\pi$. c) Inhomogeneous conventional quantum walks can possess $0$ and $\pi$ energy bound states,
even though the winding numbers associated with the both sides of the boundary are zero. Such existence of 
two flavors of topologically protected bound states is a unique feature of periodically driven systems and quantum walks. 
As is shown in \secref{sec:topological_invariant}, each bound state at $0$ or $\pi$ energy is associated with 
a topological number $\pm 1$. For a given system, a sum of these topological numbers $Q^0$ for $E=0$ and 
$Q^\pi$ for $E=\pi$ is a conserved quantity that cannot change its value unless the gap of the system closes. }
\label{fig:spectrum_zero_pi}
\end{center}
\end{figure}
There is a more direct way to confirm such robustness by simply looking at the spectrum. 
In \secref{sec:reflecting}, we found that the energies of the topological bound states is always either $E=0$ or $E=\pi$. 
Because the spectrum of the bulk (or the spectrum of the system without boundaries) is gapped, 
the total spectrum of the system studied in the previous section look as in \fref{fig:spectrum_zero_pi} a) and b), 
where a single localized state sits at $E= 0$ or $E=\pi$ 
and the bands of states away from $E=0$ or $E=\pi$ correspond to extended states in the bulk. 

Now we argue that the energy of the bound state sitting at the energy $E=0$ or $E=\pi$ cannot be 
changed by a continuous change of Hamiltonian which preserves the sublattice (chiral) symmetry. 
First of all,  the presence of sublattice (chiral) symmetry implies that 
the states with energy $E$ and $-E$ have to come in pairs.
In order to shift the energy of the bound state at $E=0$ by a small amount $\epsilon$, then, 
it is necessary to create two states at energy $\pm \epsilon$. However, 
since a single state cannot be split into two, this is impossible, and 
the energy of the single state initially at $E=0 (\pi)$ is pinned at $E=0 (\pi)$. 
According to this argument, 
the only way to remove such a zero (or $\pi$) energy bound state is to change the Hamiltonian until the bulk energy
bands close the gap so that bulk states mix with the the boundary state.

The structure of the spectrum illustrated in  \fref{fig:spectrum_zero_pi} a) and b) is generic. 
All the topological bound states exist at $E=0$ or $E=\pi$. 
It is possible to assign topological numbers to these bound states, which give yet another 
understanding of the topological protection of the bound states. 
These topological numbers are different from the winding numbers we assigned to the quantum walk 
band structures. These topological numbers are now assigned to the bound states themselves. 
The following consideration shows that 
the topological classification of the quantum walks with sublattice (chiral) symmetry is $Z \times Z$, which means 
any integer numbers of $E=0$ and $E=\pi$ energy states are topologically protected. 

In the following, we consider general one dimensional systems with sublattice (chiral) symmetry,
as in the case of quantum walks. 
Here we consider the bound states at energy $0$. Analogous arguments can be 
 applied to the bound states at $\pi$. 
 Suppose that there is $N_0$ number of degenerate bound states with energy $0$. 
 We label these states by $\ket{\phi_{\alpha'}^{0}}$ with $\alpha' =1 \cdots N_{0}$. 
  Let the sublattice (chiral) symmetry operator be given by $\Gamma$. 
  Sublattice (chiral) symmetry implies that we have the anti-commutation relation 
  between $\Gamma$ and Hamiltonian, $H$,  such that $\{ \Gamma, H\}=0$. 
  As a consequence, $\Gamma^2$ commutes with $H$.  
 When there is no conserved quantity associated with $\Gamma^2$, it is possible to 
 choose the phase of $\Gamma$ such that $\Gamma^2=1$. For example, in the case of 
 quantum walks, we choose $\Gamma = ie^{i\vec{A} \cdot \vec{\sigma} \pi/2}$. 
Because of the sublattice (chiral) symmetry, we know that $\Gamma \ket{\phi_{\alpha'}^{0}}$ is also 
 an eigenstate of $H$ with $E=0$, so $\Gamma$ represents a rotation within the subspace of zero energy states, 
 $\{ \ket{\phi_{\alpha'}^{0}} \}$. Then we can choose the basis of zero energy states such that they become 
 eigenstates of $\Gamma$. We denote 
 the zero energy states in this basis as  $\{\ket{\psi_{\alpha}^{0}}\}$ and their eigenvalues under $\Gamma$ as 
 $\{Q^0_{\alpha}\}$. 
 Since $\Gamma^2=1$, $Q^0_{\alpha}$ is either $\pm 1$. 

We now show that 
 the sum of eigenvalues, $Q^{0} \equiv \sum_\alpha Q^0_\alpha$, represents the topological invariant associated with 
 zero energy bound states. 
We define the integer number $Q^{0}$ for zero energy bound states
 and $Q^{\pi}$ for $\pi$ energy bound states constructed in an analogous fashion, as 
 \begin{eqnarray}
 Q^{0} = \sum_{\alpha} \bra{\psi_{\alpha}^0} \Gamma \ket{\psi_{\alpha}^0} \\
  Q^{\pi} = \sum_{\alpha} \bra{\psi_{\alpha}^{\pi}} \Gamma \ket{\psi_{\alpha}^{\pi}}
  \end{eqnarray}
  where $\{ \ket{\psi_{\alpha}^{\pi}} \}$ are the $\pi$ energy bound states. 
  
  In order to show that these quantities are indeed topological invariants, we show that perturbations of 
  the Hamiltonian which preserves the sublattice (chiral) symmetry cannot mix two states 
  both at the zero ($\pi$) energy with the same eigenvalues of $\Gamma$. 
This implies that such perturbations do not lift the energies of these states away from $0$ or $\pi$.
Thus, $Q^{0}$ ($Q^{\pi}$) number of bound states at energy $E=0 (E=\pi)$ cannot change 
under small deformations of the Hamiltonian. 
  
  Let $H'$ be a perturbation to the system such that  $\{ \Gamma, H'\}=0$. Now we evaluate the matrix 
  element of $\{ \Gamma, H'\}=0$ in the $0$($\pi$) energy states. The result is 
  \begin{eqnarray*}
  0 &=&  \bra{\psi_{\alpha}^0} \{ \Gamma, H'\} \ket{\psi_{\beta}^0} \\
  &=&\left\{ \begin{array}{ll} 
  2  \bra{\psi_{\alpha}^0} H'\ket{\psi_{\beta}^0} \quad \textrm{for} \quad Q^{0}_{\alpha} = Q^{0}_{\beta}  \\
  \bra{\psi_{\alpha}^0} H'\ket{\psi_{\beta}^0} -  \bra{\psi_{\alpha}^0} H'\ket{\psi_{\beta}^0}  =0
  \quad \textrm{for} \quad Q^{0}_{\alpha} \neq Q^{0}_{\beta} \end{array} \right.
  \end{eqnarray*}
  This calculation shows that bound states with the same eigenvalues $Q^{0}_{\alpha}$ 
  cannot mix. On the other hand, the same calculation does not give any constraint on the mixing of 
  states with different eigenvalues $Q^{0}_{\alpha}$. 
Because one can break up any finite change of the Hamiltonian into successive
  changes of small perturbations, one can repeat this argument
  and show that the values $Q^{0}$ and $Q^{\pi}$ cannot change unless 
  the bound states at $0$ and $\pi$ energies mix with the bulk states. 
  
  \subsection{Breaking of sublattice (chiral) symmetry}
  Here we briefly comment on the perturbations of the Hamiltonian that breaks sublattice (chiral) symmetry. 
  In the previous section, we give the argument that zero or $\pi$ energy bound states are protected as long as 
  the bands in the spectrum do not close the gap. 
  Now one can ask what happens if we consider the perturbations of Hamiltonians
  that break sublattice (chiral) symmetry. Since no topological structure can be defined in the absence
  of sublattice (chiral) symmetry, there is no longer 
  topological protection of the bound states. Yet, the statement that bound states cannot disappear until they
  mix with the bulk states remains true. Therefore, if we perturb the system that possesses topological bound states by 
  adding small perturbations that break sublattice (chiral) symmetry, the existence of the bound states is still
  protected by the existence of the gap. However, the bulk gap does not have to close to remove such a bound state;
  now the energy of the bound state can take any value in the absence of sublattice (chiral) symmetry and the state
  can be lifted away from $0$ or $\pi$ energy. 
  
  \subsection{$0$ and $\pi$ energy bound states ;  topological phenomena unique to periodically driven systems}
  \label{sec:zeropi}
  Zero-energy bound state in one dimensional system with sublattice (chiral) symmetry has been known 
  for almost 30 years, and their existence was first predicted by Su-Schrieffer-Heeger model of polyacetylene
  and Jackiw-Rebbi model\cite{Su1979, Jackiw1979, Jackiw1981}. 
  On the other hand, we saw in previous sections that quantum walks have 
  two topologically protected bound states; $0$ and $\pi$ energy states. The appearance of $\pi$ energy states
  is the result of the periodicity of quasi-energy. In return, quasi-energy is a property of periodically driven systems, 
  and thus such appearance of two flavors of topologically protected bound states is a unique phenomenon to 
  driven systems, which cannot occur in non-driven systems. 
  
  In previous examples, only one of these states, $0$ and $\pi$ energy bound states, appears in a single system. 
  From the argument given in \secref{sec:bound_state}, which of $E=0$ and $E=\pi$ 
  bound states appears across the boundary of two topological phases 
  is determined by whether the band gap closes at quasi-energy $E=0$ or $E=\pi$. 
  For example, consider the creation of a boundary between two different topological phases 
  in the inhomogeneous split-step quantum walk, as we considered
  in  \secref{sec:bound_state}. By choosing $\theta_{2-}$ and $\theta_{2+}$ appropriately, one can either make the boundary 
  between the two phases to be gapless at either $E=0$ or $E=\pi$, as one can see from the phase diagram in \fref{fig:split_step}. When the boundary closes the gap at $E=0 (\pi)$, the bound state appears at $E=0 (\pi)$
    as is depicted in \fref{fig:spectrum_zero_pi} a) and b). 
  
 Now consider the creation of the boundary between the phases with {\it the same winding numbers} 
 by setting $\theta_{1} =0$. This quantum walk whose evolution operator is given by $U= T_{\downarrow} R_{y}(\theta_{2}) T_{\uparrow}$ is nothing but the conventional quantum walk described in \secref{sec:intro} with initial time shifted. 
 This time-shifted quantum walk realizes only a single phase with $W=0$\cite{footnote4}. If we set 
 $\theta_{2-} $ to be $ - 2\pi < \theta_{2-} < 0 $ and  $\theta_{2+} $ to be $ 0 < \theta_{2+} < 2 \pi$, 
 then the two phases corresponding to $x \rightarrow -\infty$ and $x \rightarrow$ both have $W=0$. From the point of 
 view of the winding topological number defined on the band structures, 
 one expects no topologically protected bound states to be present. 
 However, this inhomogeneous quantum walk possesses two topological bound states at quasi-energies 
 $E=0$ and $E=\pi$. 
 
 The existence of two bound states near the origin can be easily confirmed for the simple case of 
 $\theta_{2-} = -\pi$ and  $\theta_{2+} = \pi$ where the rotation $\theta_{2-}$ is applied to all the sites $ x \leq 0$ 
 and the rotation $\theta_{2+}$ is applied to sites $0 < x$. These rotations act on the spins as 
 \begin{eqnarray*}
 \theta_{2-} = -\pi &:& \ket{\uparrow} \rightarrow - \ket{\downarrow},  \ket{\downarrow} \rightarrow \ket{\uparrow} \\
  \theta_{2+} = \pi &:& \ket{\uparrow} \rightarrow  \ket{\downarrow},  \ket{\downarrow} \rightarrow  - \ket{\uparrow} 
 \end{eqnarray*}
 Now we consider the evolution of the particle after one step for a particle at site $x=1$ with spin either up or down. 
 This state evolves as 
  \begin{eqnarray*}
&&\ket{0} \otimes \ket{\uparrow} 
\stackrel{T_{\uparrow}}{\rightarrow } \ket{1} \otimes \ket{\uparrow} 
\stackrel{R}{\rightarrow } \ket{1} \otimes \ket{\downarrow} 
\stackrel{T_{\downarrow}}{\rightarrow } \ket{0} \otimes  \ket{\downarrow}  \\
&&\ket{0} \otimes \ket{\downarrow} 
\stackrel{T_{\uparrow}}{\rightarrow } \ket{0} \otimes \ket{\downarrow} 
\stackrel{R}{\rightarrow } \ket{0} \otimes  \ket{\uparrow} 
\stackrel{T_{\downarrow}}{\rightarrow } \ket{0} \otimes   \ket{\uparrow} 
 \end{eqnarray*}
 Thus it is clear that 
 $\ket{\psi_{E=0}} = \ket{0} \otimes \frac{1}{\sqrt{2}} \left( \ket{\uparrow} + \ket{\downarrow} \right)$ is an eigenstate 
 of the one-step evolution operator with quasi-energy $0$ and 
 $\ket{\psi_{E=\pi}} = \ket{0} \otimes \frac{1}{\sqrt{2}} \left( \ket{\uparrow} - \ket{\downarrow} \right)$ 
 has quasi-energy $\pi$. It is straightforward to
 check that any other states in this system has energy $E=\pm \pi/2$. 
 Because this system possesses a single bound state at energy $E=0$ and $E=\pi$, these states cannot be removed 
 from these states until the gap of the bulk states closes, and thus these states must be present for any 
 parameter values $ - 2\pi < \theta_{2-} < 0 $ and $ 0 < \theta_{2+} < 2 \pi$. 
 The general spectrum of such inhomogeneous quantum walks is illustrated in \fref{fig:spectrum_zero_pi} c). 
 As we noted in \secref{sec:topological_invariant}, these bound states are associated with topological numbers.
 In this walk, since $\theta_{1}=0$, the sublattice (chiral) symmetry is given by the operator $\Gamma= \sigma_{x}$. 
 Then the topological number associated with the bound state $\ket{\psi_{E=0}}$ is nothing but the 
 eigenvalue of $\Gamma$, so $\ket{\psi_{E=0}}$ has $Q^{0} =1$ and $\ket{\psi_{E=\pi}}$ has $Q^{\pi} =-1$. 
 
 Such coexistence of $0$ and $\pi$ energy bound state can also be observed in inhomogeneous split-step 
 quantum walks, where the two phases on the left and on the right are separated by two gapless phases 
 where one closes the gap at $E=0$ and the other at $E=\pi$. 

The winding number is the comprehensive topological description 
of static Hamiltonians, but quantum walks are periodically 
driven systems. More completely, periodically driven systems should be described by
 the evolution operator over one period, and the topological numbers for such systems should be written in terms of 
 the evolution operator and not in terms of the static effective Hamiltonian. 
 Thus the topological classification of quantum walks 
 is not given by ${\bf Z}$ as for the winding numbers, but in fact given by ${\bf Z} \times {\bf Z}$ as we have seen
 in the topological invariants of bound states in \secref{sec:topological_invariant}. More detailed analysis of 
 the difference of topological classification between static systems and periodically driven systems is given in 
[\onlinecite{Kitagawa2010b}].

   \begin{figure}[t]
\begin{center}
\includegraphics[width = 8.5cm]{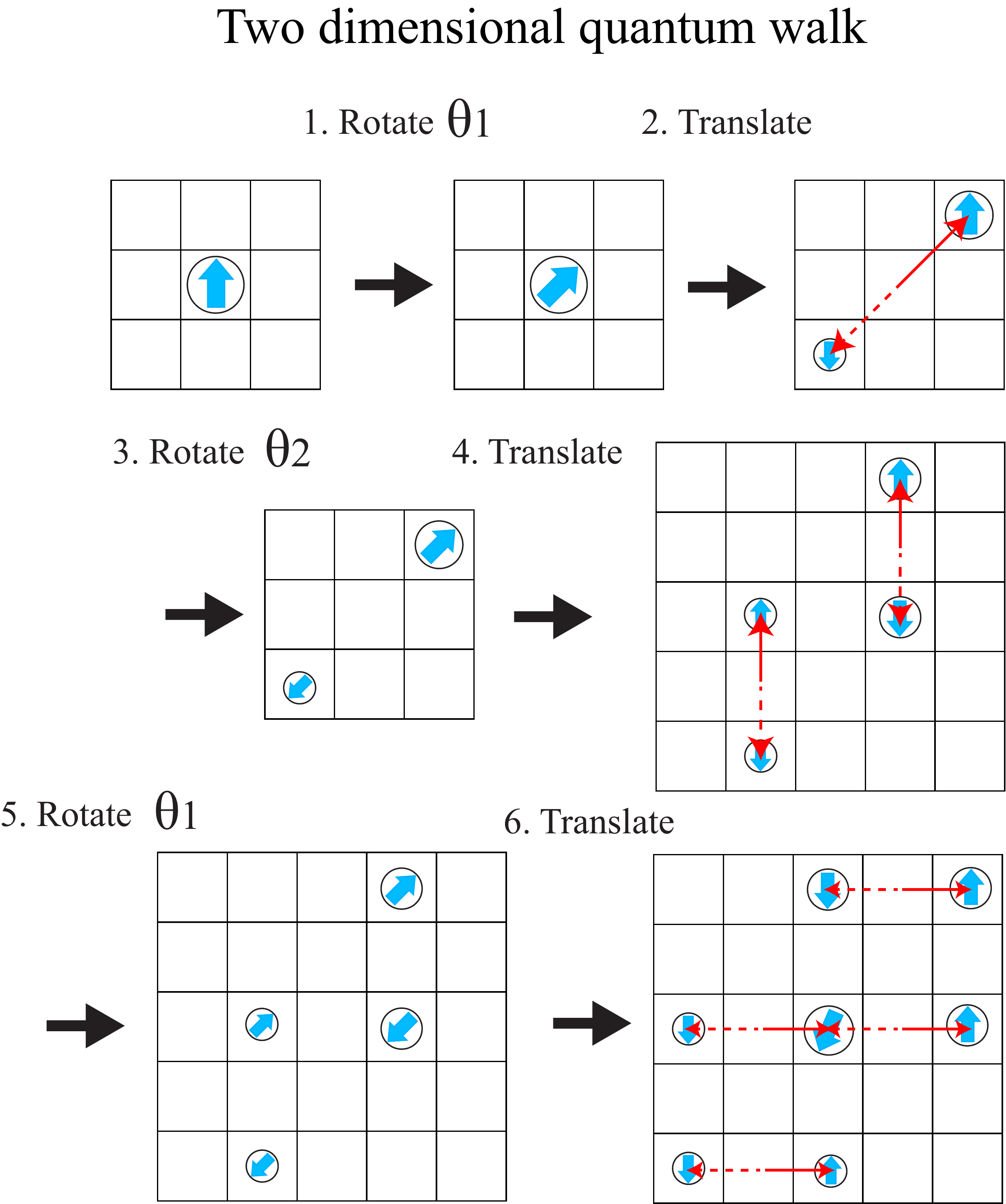}
\caption{Protocol of two dimensional quantum walk considered in \secref{sec:2D_quantum_walk}.
The quantum walk is defined for a single spin $1/2$ particle in two dimensional lattice. 
The protocol consists of $6$ operations. In the figure, the spin-dependent translation is denoted by 
red arrows, where solid arrow is the movement for spin up and dotted arrow is for spin down. 
The protocols are
 1. spin rotation around $y$ axis by angle $\theta_{1}$; 
2. spin-dependent translation where spin up is move to right and up by one lattice site, and spin down is moved 
to left and down; 3. spin rotation around $y$ axis by angle $\theta_{2}$; 4. spin-dependent translation where 
spin up is moved to up, and spin down to down; 5.  spin rotation around $y$ axis by the same rotation angle as the 
first rotation $\theta_{1}$; 6. spin-dependent translation where spin up is move to right and down to left. 
Each step of quantum walk takes a particle from even (odd) coordinate to even (odd) coordinate, so the lattice constant
of the effective Hamiltonian is $2$. Thus, the first Brillouin zone is $-\pi/2 \leq k_{x} \leq \pi/2$ and  $-\pi/2 \leq k_{y} \leq \pi/2$. }
\label{fig:2D_scheme}
\end{center}
\end{figure}
\section{Quantum walks in two dimension} \label{sec:2DQW}
\subsection{effective Hamiltonian and Chern number} \label{sec:2D_quantum_walk}
In the previous sections, we illustrated the ideas of quantum walks and topological phases realized in
these systems in the simplest setting, one dimensional quantum walks with two internal degrees of freedom. 
However, the idea of topological phases is much more general, and it is possible to extend the 
quantum walk protocols to study different topological phases in different dimensions.
In this section, we illustrate the idea by describing the two dimensional quantum walks and demonstrating 
that this quantum walk realizes topological phases with Chern numbers, the phases that are responsible for
integer quantum Hall effects that we explained in \secref{sec:topological_phases}.

We consider the quantum walk of spin $1/2$ particle on a square lattice. 
In the literature, quantum walks in dimensions larger than $1$ are defined for larger number of 
internal degrees of freedom, but the quantum walk defined here is simpler and easier to realize in experiments. 
The quantum walk consists of three rotations and three translations, implemented in alternative fashion 
(see \fref{fig:2D_scheme}); 
\begin{enumerate}
\item Rotation of the spin around $y$ axis by
angle $\theta_{1}$,  given by $R_{y}(\theta_{1})= e^{-i\theta_{1} \sigma_{y}/2}$. 
\item Translation of the spin $\uparrow$ one lattice to the right and up, and 
translation of the spin $\downarrow$ one lattice to the left and down. Explicitly, \\ 
$T_{1} = \sum_{x, y} \ket{x+1, y+1} \bra{x, y} \otimes \ket{\uparrow}\bra{\uparrow} \\ + 
\ket{x-1, y-1} \bra{x, y} \otimes \ket{\downarrow}\bra{\downarrow}$. 
\item Rotation of the spin around $y$ axis by
angle $\theta_{2}$,  given by $R_{y}(\theta_{2})= e^{-i\theta_{2} \sigma_{y}/2}$. 
\item Translation of the spin $\uparrow$ by one lattice to up, and 
translation of the spin $\downarrow$ by one lattice to down. Explicitly, \\ 
$T_{2} = \sum_{x, y} \ket{x, y+1} \bra{x, y} \otimes \ket{\uparrow}\bra{\uparrow} \\ + 
\ket{x, y-1} \bra{x, y} \otimes \ket{\downarrow}\bra{\downarrow}$. 
\item Rotation of the spin around $y$ axis by the same 
angle as the first rotation $\theta_{1}$,  given by $R_{y}(\theta_{1})= e^{-i\theta_{1} \sigma_{y}/2}$. 
\item Translation of the spin $\uparrow$ by one lattice to right, and 
translation of the spin $\downarrow$ by one lattice to left. Explicitly, \\ 
$T_{3} = \sum_{x, y} \ket{x+1, y} \bra{x, y} \otimes \ket{\uparrow}\bra{\uparrow} \\ + 
\ket{x-1, y} \bra{x, y} \otimes \ket{\downarrow}\bra{\downarrow}$. 
\end{enumerate}
Note that in this quantum walk, the particle after one step of quantum walk moves from even (odd) coordinate
sites to even (odd) coordinate sites as one can see from \fref{fig:2D_scheme}. 
Thus the effective Hamiltonian of the quantum walk has the lattice constant
equal to $2$. Therefore, for translationally invariant quantum walks, the first Brillouin zone is given by
$-\pi/2 \leq k_{x} \leq \pi/2$ and $-\pi/2 \leq k_{y} \leq \pi/2$. 
The evolution of the particle distribution in this walk can be studied in a similar fashion as 
the one dimensional analogue, and in particular, the asymptotic distribution is obtained through 
the formalism developed in \secref{sec:asymptotic}. 

 \begin{figure}[b]
\begin{center}
\includegraphics[width = 8.5cm]{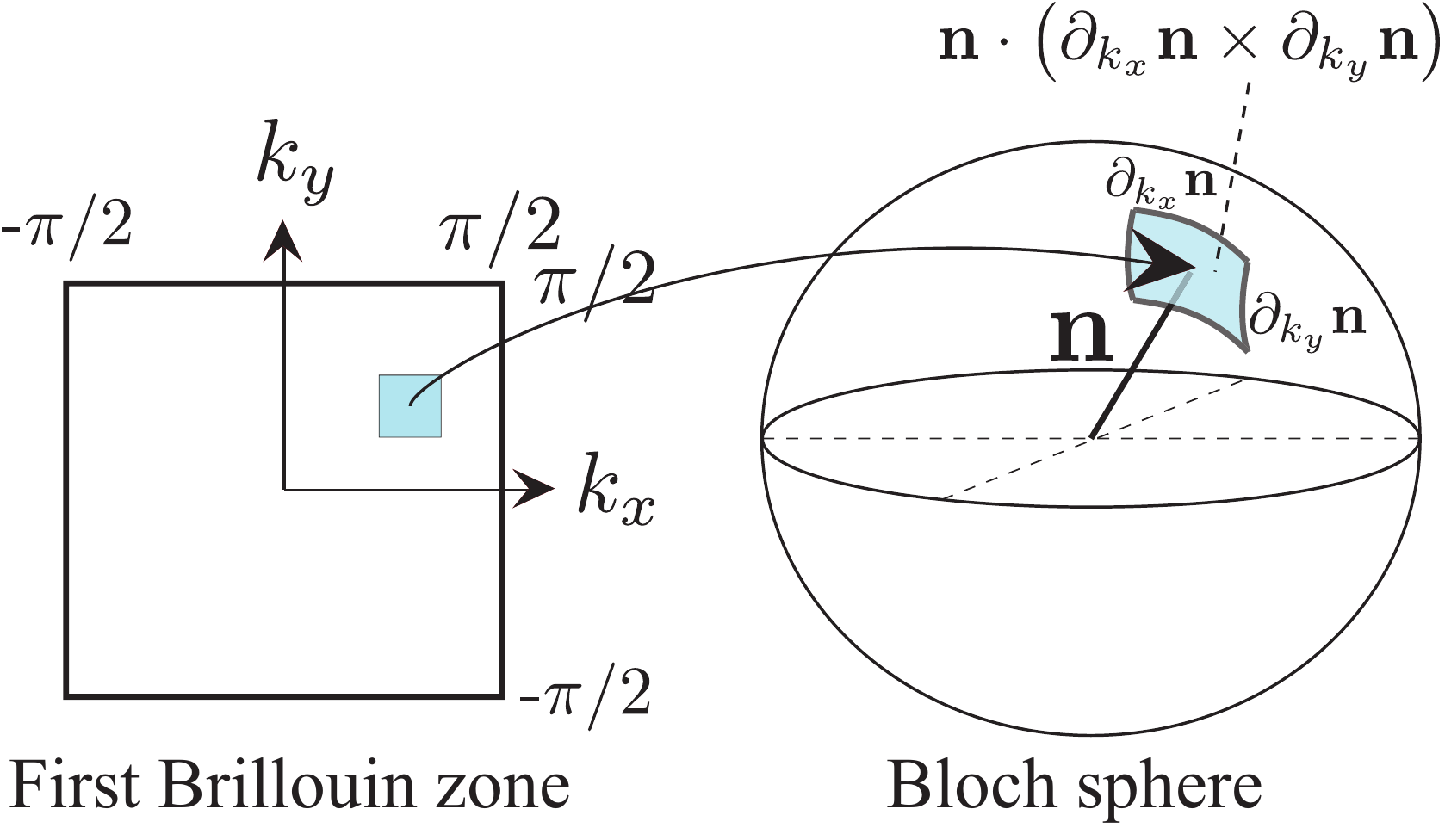}
\caption{Illustration of Chern number of two dimensional systems with two bands. The eigenstate for each quasi-momentum $\vec{k}$ is a superposition of spin up and down, and can be represented as a point on Bloch sphere, given by
$\vec{n}(\vec{k})$. Thus $\vec{n}(\vec{k})$ represents a map from the first Brillouin zone to Bloch sphere. In order to obtain
a topological number for this system, we consider the area mapped by $\vec{n}(\vec{k})$ from the first Brillouin zone to 
Bloch sphere. Due to the periodic boundary condition of the first Brillouin zone, which is a torus, such map has to 
wrap around the Bloch sphere by integer number of times. This integer is what is called a Chern number, and represents
the topological number associated with the system. The formal expression of the Chern number is then obtained by
calculating the area covered by the gap $\vec{n}(\vec{k})$, which can be calculated in the way illustrated in this picture and results in the expression \eqnref{chern_number} }
\label{fig:chern_number}
\end{center}
\end{figure}

 \begin{figure*}[th]
\begin{center}
\includegraphics[width = 17cm]{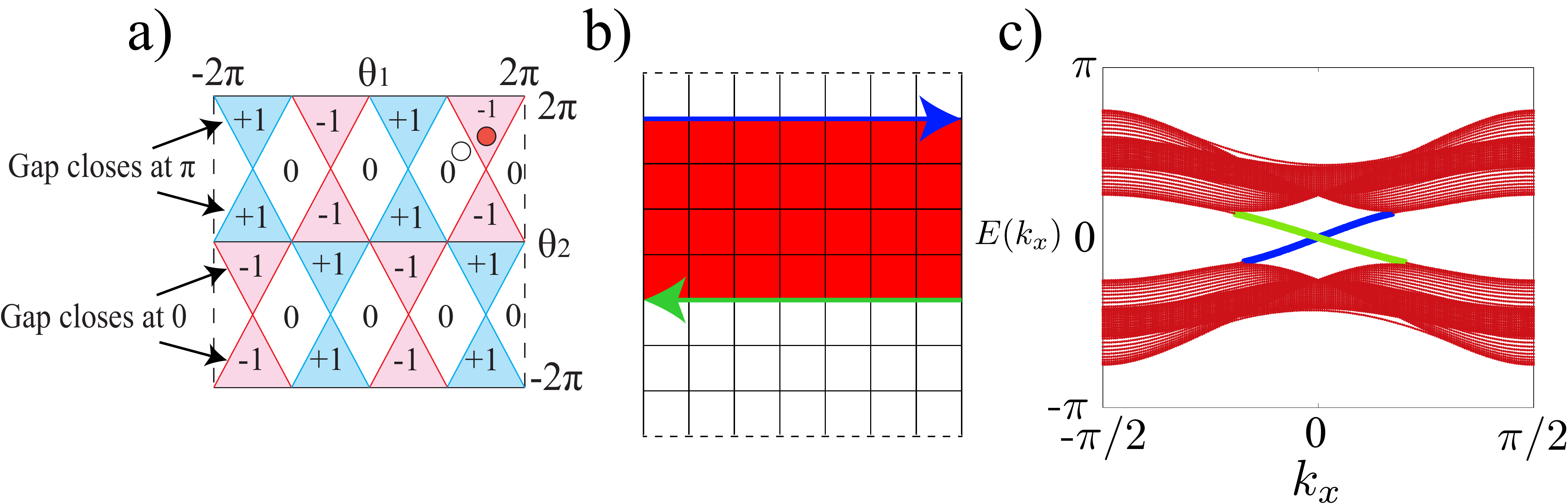}
\caption{a) Phase diagram of the two dimensional quantum walk. Each phase is characterized by a Chern number of a lower band with quasi-energy $- E(\vec{k})$, which can take values $0, \pm 1$ in this quantum walk. 
The Chern number can change only when the system crosses gapless phases, and the lines of gapless phases are 
indicated by the red and blue line in the diagram. b) physical manifestation of Chern numbers appears at the boundary
between regions that belong to phases with different Chern numbers. Here we illustrate the inhomogeneous quantum walk, where the quantum walk in the central region (colored as red) corresponds to $\theta_{1+} = 7 \pi/6$ and $\theta_{2+} = 7 \pi/6$ (red dot in a)), whereas in the other half (colored as white), the quantum walk corresponds to $\theta_{1-} = 3 \pi/2$ and $\theta_{2-} = 3 \pi/2$ (white dot in a)). In the simulation, we took the periodic boundary condition for both $x$ and $y$ direction, and the system size
is $100 \times 100$.  c) Quasi-energy spectrum of the inhomogeneous quantum walks illustrated in b). The states colored
as red are bulk states, and the states that go from the lower band to the upper band are the 
unidirectionally propagating edge states that are localized
near the boundary of the two phases. States colored 
as blue are the states that run along the upper edge and those colored as 
green are the states that run along the lower edge, as illustrated in b).  }
\label{fig:2D_phase_diagram}
\end{center}
\end{figure*}

In order to study the topological properties of this quantum walk, we consider the effective Hamiltonian 
of the quantum walk. As we detail in Appendix \ref{appendix:2D_spectrum}, the effective Hamiltonian takes the form 
\begin{equation}
H_{\textrm{eff}} = \sum_{\vec{k}}  E(\vec{k}) \vec{n} (\vec{k}) \cdot \sigma \otimes \ket{\vec{k}}\bra{\vec{k}}
\end{equation} 
The spectrum $E(\vec{k}) \vec{n} (\vec{k})$ is determined by the equation 
\begin{eqnarray} \label{2D_energy}
 \cos \left( E(\vec{k}) \right) &=& \cos( k_{x}) \cos( k_{x}+ 2k_{y}) \cos(\theta_{1}) \cos(\theta_{2}/2)  \nonumber \\ 
&&- \sin( k_{x}) \sin( k_{x}+ 2k_{y}) \cos(\theta_{2}/2)  \nonumber \\
&& - \cos^2( k_{x}) \sin( \theta_{1}) \sin(\theta_{2}/2) 
\end{eqnarray}

The topological structure of two dimensional system appears in $\vec{n} (\vec{k})$ as in the case
of one dimensional quantum walk. Since now we have two dimensional Brillouin zone, the 
function $\vec{n} (\vec{k})$ is a map from two dimensional torus to Bloch sphere, see \fref{fig:chern_number}. 
A small area on the torus is mapped to the small area on the Bloch sphere. If one maps the total area
of the torus onto the Bloch sphere, the map necessarily wraps around the sphere an integer number 
of times due to the periodic boundary condition of the torus. Thus, if one calculates the total area
covered by the map $\vec{n} (\vec{k})$, the value is $4\pi n$ where $n$ is an integer. This integer 
is so-called Chern number, which is responsible for integer quantum Hall effect in two dimensional
electron gas. Explicitly, the Chern number can be expressed in terms of $\vec{n} (\vec{k})$ as 
\begin{equation} \label{chern_number}
C = \frac{1}{4\pi} \int_{\textrm{FBZ}} d\vec{k} \, \, \vec{n} \cdot \left( \partial_{k_{x}} \vec{n} \times \partial_{k_{y}} \vec{n} \right)
\end{equation} 
As opposed to the winding number of one dimensional quantum walk, this topological number does not 
rely on any symmetry of the system, and thus can exist in the absence of any symmetry. 

Conventionally, Chern number is associated with each band of Hamiltonian. The definition of Chern number above
gives the Chern number of "lower" band with quasi-energy of $-E(\vec{k})$ in \eqnref{2D_energy}, 
and the Chern number of upper band is given by simply $- C$ so that
the Chern numbers of all the bands sum to zero. More generally, if the wavefunction of a band at a given quasi-momentum
$\vec{k}$ is given by $\ket{\psi(\vec{k})} = e^{i\vec{r} \cdot \vec{k}} \ket{\phi(\vec{k})}$, 
where $\ket{\phi(\vec{k})}$ is the periodic part of the Bloch wave function, then the Chern number associated with the band 
is given by the famous TKNN formula\cite{Thouless1982}; 
\begin{eqnarray} \label{TKNN}
C &=& \frac{1}{2\pi} \int_{\textrm{FBZ}} d \vec{k} \left[ \partial_{k_{x}} A_{k_{y}} -  \partial_{k_{y}} A_{k_{x}} \right] \\
(A_{k_{x}} ,  A_{k_{y}} ) & = &  ( i \bra{ \phi(\vec{k})} \partial_{k_{x}} \ket{\phi(\vec{k})},  i \bra{ \phi(\vec{k})} \partial_{k_{y}} \ket{\phi(\vec{k})}) \nonumber
\end{eqnarray}
This TKNN formula calculates the Berry phase of an electron as it goes around the first Brillouin zone. 
This expression of Chern number reduces to \eqnref{chern_number} in the case of systems with
two bands.

One can calculate the Chern numbers for the two dimensional quantum walk described above 
for various values of $\theta_{1}$ and $\theta_{2}$. The phase diagram is plotted in \fref{fig:2D_phase_diagram} a). 
A convenient way to obtain such phase diagram is to first obtain the lines of gapless phases.
Since the topological number can only change its value across gapless phase, it is only necessary to 
compute the Chern number at a single point of the gapped region and any gapped phase that is continuously 
connected with that point without crossing gapless phase must have the same topological number as that point. 
We detail the calculation to identify gapless phases for this two dimensional quantum walk in the 
Appendix \ref{appendix:2D_gapless}. 

A physical manifestation of topological phases with Chern number is, as is the case for one dimensional quantum walk,
boundary states across the regions in which two different topological phases are realized. The existence of such bound states can be understood according to the argument given in \secref{sec:bound_state}; the band structures away from 
the boundary are gapped, but the band gap has to close near the boundary in order for the topological number, 
Chern number, to change its value. Thus there is generically states in the gap of the bulk systems and these states 
are necessarily localized near the boundary. 

These bound states that appear in systems with non-zero Chern number are known to 
propagate in unidirectional fashion without any backscattering. It is possible to confirm 
the existence of such unidirectional edge states
by considering inhomogeneous quantum walk where the particle at sites with $0 \leq y$ evolves according to 
the two dimensional quantum walk with rotation angles $\theta_{1+}$ and $\theta_{2+}$, and the particle 
at  sites with $y < 0$ evolves according to 
the two dimensional quantum walk with rotation angles $\theta_{1-}$ and $\theta_{2-}$. 
Such inhomogeneous quantum walk is illustrated in \fref{fig:2D_phase_diagram} b). 
If the Chern number of 
the phases corresponding to $\theta_{1+}$ and $\theta_{2+}$ and $\theta_{1-}$ and $\theta_{2-}$ are different, 
unidirectional edge states are expected to appear along $y=0$. 

Since these edge states exist, just like $0$ and $\pi$ energy states of one dimensional quantum walk, in the gap
of the bulk states, it is easy to identify the existence of these states by numerically solving for the quasi-energy spectrum. 
In \fref{fig:2D_phase_diagram} c), we provided the plot of quasi-energy spectrum for a torus geometry with periodic boundary condition on 
both $x$ and $y$ direction. The system size is taken to be $100 \times 100$. In the upper half of the 
system between $0 \leq y < 50$, 
we implemented the quantum walk with $\theta_{1+} = 7 \pi/6$ and $\theta_{2+} = 7 \pi/6$, whereas
in the lower half $-50 \leq y <0$, the quantum walk corresponds to 
$\theta_{1-} = 3 \pi/2$ and $\theta_{2-} = 3 \pi/2$. 
Note that there are two boundary in this system, corresponding to the lower edge at $y=0$ and upper edge
at $y =50$. In the spectrum, there are clearly 
two edge states, colored as red and green, which run along the upper and lower edge, respectively. 
These chiral edge modes are the signature of Chern numbers in two dimensional quantum walk. 

\subsection{Unidirectionally propagating modes in quantum walks without Chern numbers} \label{sec: simple_2D}
In the case of one dimensional quantum walk, we found the existence of two bound states at quasi-energy
 $0$ and $\pi$ near the boundary of the phases with zero winding number.
 This existence of two flavors of topologically protected bound states represented 
 a phenomenon unique to periodically driven systems that do not exist in static systems, and thus 
 the existence is not captured by the winding number of the effective Hamiltonian. 
 
 In a similar fashion, it is possible to have 
 unidirectionally propagating modes across the boundary of the regions where quantum walks 
 in each region have no Chern numbers associated with the phases. 
 In the following, we show that such chiral propagating modes exist 
 for even simpler version of two dimensional quantum walk protocols. 
 
  \begin{figure}[t]
\begin{center}
\includegraphics[width = 8.5cm]{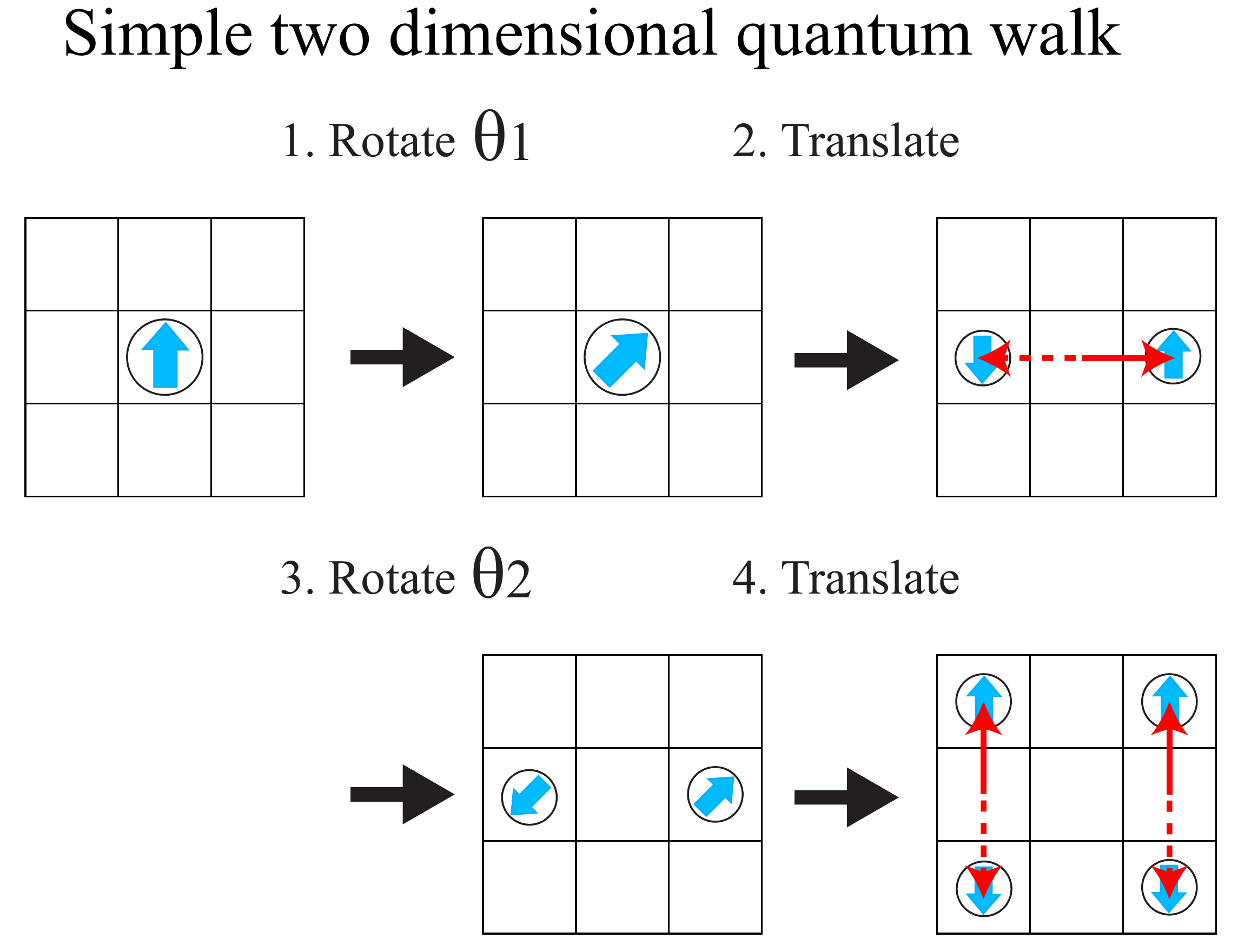}
\caption{Protocol for the simple two dimensional quantum walk studied in \secref{sec: simple_2D}. 
In this protocol, only four operations are applied during one step of quantum wallk. As before, 
the spin-dependent translation is indicated by red arrows, where solid arrow is for spin up and dotted arrow
is for spin down.  The explicit protocol is; 
1. spin rotation around $y$ axis by angle $\theta_{1}$; 2. spin-dependent translation where 
spin up is move to right, and spin down to left; 3. spin rotation around $y$ axis by angle $\theta_{2}$; 
4. spin-dependent translation where spin up is moved to up, and spin down to down. }
\label{fig:2D_simple_scheme}
\end{center}
\end{figure}
 Here we consider the following simple 
 two dimensional quantum walk with two rotations and two spin dependent translations,
 see \fref{fig:2D_simple_scheme};
 \begin{enumerate}
\item Rotation of the spin around $y$ axis by
angle $\theta_{1}$,  given by $R_{y}(\theta_{1})= e^{-i\theta_{1} \sigma_{y}/2}$. 
\item Translation of the spin $\uparrow$ one lattice to the right, and 
translation of the spin $\downarrow$ one lattice to the left. Explicitly, \\ 
$T_{1} = \sum_{x, y} \ket{x+1, y} \bra{x, y} \otimes \ket{\uparrow}\bra{\uparrow} \\ + 
\ket{x-1, y} \bra{x, y} \otimes \ket{\downarrow}\bra{\downarrow}$. 
\item Rotation of the spin around $y$ axis by
angle $\theta_{2}$,  given by $R_{y}(\theta_{2})= e^{-i\theta_{2} \sigma_{y}/2}$. 
\item Translation of the spin $\uparrow$ one lattice to the up, and 
translation of the spin $\downarrow$ one lattice to the down. Explicitly, \\ 
$T_{2} = \sum_{x, y} \ket{x, y+1} \bra{x, y} \otimes \ket{\uparrow}\bra{\uparrow} \\ + 
\ket{x, y-1} \bra{x, y} \otimes \ket{\downarrow}\bra{\downarrow}$. 
\end{enumerate}

  \begin{figure*}[t]
\begin{center}
\includegraphics[width = 13cm]{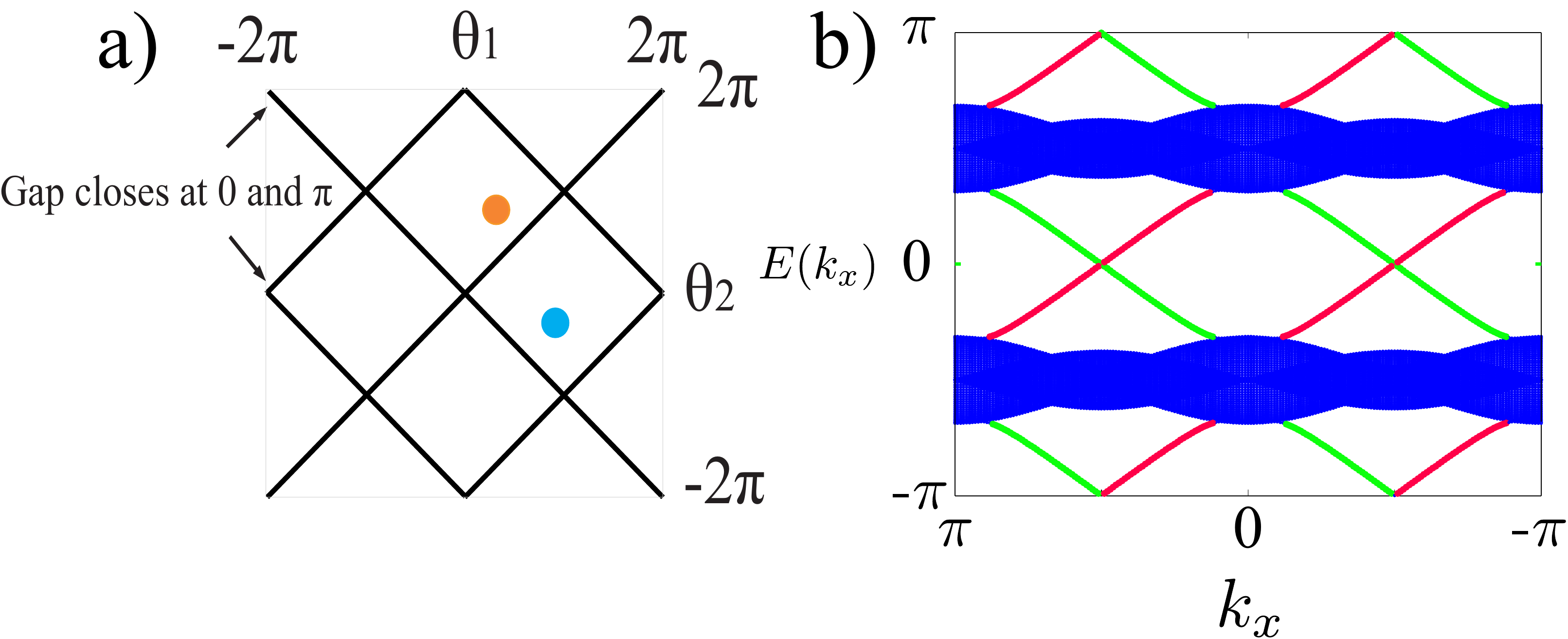}
\caption{a) Phase diagram of simple two dimensional quantum walk studied in \secref{sec: simple_2D}. 
The Chern number of the quantum walk is everywhere zero. Yet, there are topologically protected 
unidirectionally propagating modes in the inhomogeneous quantum walk. 
The existence of such unidirectional edge states are shown through analytical calculations for
special values of quantum walks in the text. 
b) quasi-energy spectrum of the inhomogeneous quantum walks, where the two quantum walk protocols 
corresponding to the two regions are indicated as orange and blue dot in a). The bulk states are colored as 
blue, and unidirectionally propagating states are colored as green and red. Green states propagate along the 
lower edge of the boundary and red states along the upper edge. For a given edge, say, lower edge, the 
edge states have non-zero energy winding as $k_{x}$ goes from $-\pi$ to $\pi$, and thus, such edge states 
cannot be removed under the continuous change of quantum walk protocols
unless the bulk gap closes, where the upper edge and lower edge are allowed to mix.  }
\label{fig:simple_2D_phase_diagram}
\end{center}
\end{figure*}

The effective Hamiltonian of this quantum walk is again given by the form 
$H_{\textrm{eff}} = \sum_{\vec{k}}  E(\vec{k}) \vec{n} (\vec{k}) \cdot \sigma \otimes \ket{\vec{k}}\bra{\vec{k}}$.
The spectrum of this quantum walk is given 
\begin{eqnarray*}
\cos(E(\vec{k}) &=& \cos(k_{x} + k_{y}) \cos (\theta_{1}/2) \cos(\theta_{2}/2) \\
&& - \cos(k_{x}-k_{y}) \sin(\theta_{1}/2)\sin(\theta_{2}/2)
\end{eqnarray*}

This quantum walk is described by Chern number zero phase everywhere, and the phase diagram is 
given in \fref{fig:simple_2D_phase_diagram} a). All the gapless phases close their gap at both $0$ and $\pi$ energy. 
Now consider the inhomogeneous quantum walks in this protocol, where the particle is controlled by
the quantum walk protocol with rotation angles $\theta_{1+}$ and $\theta_{2+}$ at sites $0 \leq y$, 
and the protocol  at  sites with $y < 0 $ is given by the rotation angles $\theta_{1-}$ and $\theta_{2-}$. 
If we choose the angles such that the two phases are separated by a single gapless phase, 
there are in fact two unidirectionally propagating modes at the boundary.

This can be most easily confirmed for the spacial rotation angles $\theta_{1+} = 0$ and $\theta_{2+} = \pi$
and $\theta_{1-} = \pi$ and $\theta_{2-} = 0$ by simply considering the evolution for spin up and down 
for a few steps. Near the boundary, the evolution is 
\begin{eqnarray*}
&& \ket{j,0} \otimes \ket{\uparrow} 
\stackrel{R}{\rightarrow }  \ket{j,0} \otimes  \ket{\uparrow}
 \stackrel{T_{1}}{\rightarrow }  \ket{j+1,0} \otimes \ket{\uparrow} \\
&&  \stackrel{R}{\rightarrow }  \ket{j+1,0} \otimes \ket{\downarrow} 
    \stackrel{T_{2}}{\rightarrow }  \ket{j+1,-1} \otimes \ket{\downarrow}  \\
&& \ket{j,-1} \otimes \ket{\downarrow} 
\stackrel{R}{\rightarrow }  - \ket{j,-1} \otimes  \ket{\uparrow}
 \stackrel{T_{1}}{\rightarrow }  - \ket{j+1,-1} \otimes \ket{\uparrow} \\
&&  \stackrel{R}{\rightarrow }  - \ket{j+1,-1} \otimes \ket{\uparrow} 
    \stackrel{T_{2}}{\rightarrow }  -\ket{j+1,0} \otimes \ket{\downarrow}  \\
\end{eqnarray*}
Thus we see that spin up states at site $y=0$ and spin down state at site $y=-1$ both propagate
to the right during the evolution. 

By Fourier transform in $x$ coordinate, it is clear that the walk takes 
$\ket{k_{x},y = 0} \otimes \ket{\uparrow} \rightarrow  e^{ik_{x}} \ket{k_{x},y = -1} \otimes \ket{\downarrow} $ and 
$\ket{k_{x},y = -1} \otimes \ket{\downarrow} \rightarrow  - e^{ik_{x}} \ket{k_{x},y = 0} \otimes \ket{\uparrow} $.
Thus we conclude there are two unidirectionally propagating modes 
\begin{eqnarray*}
\ket{\psi_{1}} &=& \frac{1}{\sqrt{2}} \left( \ket{k_{x},y = 0} \otimes \ket{\uparrow} + i \ket{k_{x},y = -1} \otimes \ket{\downarrow} \right) \\ 
 \quad E(k_{x}) &= & k_{x} - \frac{\pi}{2} \\
\ket{\psi_{2}} &=& \frac{1}{\sqrt{2}} \left( \ket{k_{x},y = 0} \otimes \ket{\uparrow} - i \ket{k_{x},y = -1} \otimes \ket{\downarrow}  \right) \\ 
 \quad E(k_{x}) &=& k_{x} + \frac{\pi}{2} \\
\end{eqnarray*}

On the other hand, other states in the system evolve as, for $l>0$
\begin{eqnarray*}
&& \ket{j,l>0} \otimes \ket{\uparrow} 
 \stackrel{R}{\rightarrow }  \ket{j,l} \otimes  \ket{\uparrow}
 \stackrel{T_{1}}{\rightarrow }  \ket{j+1,l} \otimes \ket{\uparrow} \\ 
&&  \stackrel{R}{\rightarrow }  \ket{j+1,l} \otimes \ket{\downarrow} 
    \stackrel{T_{2}}{\rightarrow }  \ket{j+1,l-1} \otimes \ket{\downarrow}  \\
&& \stackrel{R}{\rightarrow }  \ket{j+1,l-1} \otimes \ket{\downarrow}
 \stackrel{T_{1}}{\rightarrow }   \ket{j,l-1} \otimes \ket{\uparrow} \\
 &&  \stackrel{R}{\rightarrow }  - \ket{j,l-1} \otimes \ket{\uparrow} 
    \stackrel{T_{2}}{\rightarrow }  -\ket{j,l} \otimes \ket{\uparrow}  \\
\end{eqnarray*}
Thus we conclude that the states 
$\frac{1}{2} \left( \ket{j,l} \otimes \ket{\uparrow} + i \ket{j+1,l-1} \otimes \ket{\downarrow}  \right)$ are eigenstates 
of the system with the flat quasi-energy $E = -\pi/2$ and 
$\frac{1}{2} \left( \ket{j,l} \otimes \ket{\uparrow} - i \ket{j+1,l-1} \otimes \ket{\downarrow}  \right)$
are another set of eigenstates with quasi-energy $E=\pi/2$. In a similar fashion, it is straightforward to 
show that the bulk states in $l<-1$ have energies $\pm \pi/2$. 

Notice that the edge states obtained above has a non-zero winding in the energy direction as $k_{x}$ goes
from $-\pi$ to $\pi$ {\it i.e.} the states run from quasi-energy $E = -\pi$ to $E=\pi$ as $k_{x}$ goes from 
$-\pi$ to $\pi$. it is straightforward to convince oneself, by drawing the spectrum, that 
such energy winding cannot be removed under the continuous change of quantum walk protocols unless 
the bulk gap closes. Thus, the existence of these states is guaranteed for the inhomogeneous quantum walk 
which has $\theta_{1+}$ and $\theta_{2+}$ that are continuously connected with $\theta_{1+}=0$ and $\theta_{2+}=\pi$
and $\theta_{1-}$ and $\theta_{2-}$ that are continuously connected with $\theta_{1-}=\pi $ and $\theta_{2-}=0$, 
without crossing the gapless phase. 

As an example, we plot the quasi-energy spectrum of inhomogeneous quantum walk with 
$\theta_{1+}=\pi/8$ and $\theta_{2+}=7\pi/8$ from $ 0 \leq y <50$, and $\theta_{1-}=7\pi/8$ and $\theta_{2-}=-\pi/4$
from $-50\leq y <0$, and we again take periodic boundary condition for both $x$ and $y$ direction
with system size $100 \times 100$. 
These two phases are connected with the limit 
$\theta_{1+}=0$ and $\theta_{2+}=\pi$, and $\theta_{1-}=\pi$ and $\theta_{2-}=0$ in a continuous fashion, 
as one can check. The spectrum of this system is plotted in \fref{fig:simple_2D_phase_diagram} b), 
and one can see the existence of unidirectionally
propagating modes on the two edges, colored as red and blue. 
One observes these edge modes in fact winds in the energy direction with non-trivial winding number. 
Such energy winding is in fact closely related to the phenomenon of Thouless pump\cite{Thouless1983}, and we refer
the interested readers to the detailed analysis in [\onlinecite{Kitagawa2010b}]. 

\section{Other topological phases} \label{sec:others}
Different class of topological phases exist in various symmetries and dimensions, as is classified for 
non-interacting static Hamiltonian\cite{Schnyder2008,Qi2008,Kitaev2009}. 
As we have illustrated the ideas through a few examples in this review article, it is possible to realize 
any of the topological phases in $1$ and $2$ dimension through
the variations of quantum walk protocols.
We refer the interested readers to the article [\onlinecite{Kitagawa2010}] for more complete analysis. 

Furthermore, we have given two examples of topological phenomena that are unique to periodically driven
systems; $0$ and $\pi$ energy bound states in zero winding number phases and energy winding unidirectional
edge states in zero Chern number phases\cite{Kitagawa2011, Kitagawa2010b}. 
These phenomena can also be extended to other classes of 
driven systems. Recently it has been proposed that two flavors of Majorana Fermions can be realized at 
$0$ and $\pi$ energies in cold atoms\cite{Jiang2011}, in a similar fashion as $0$ and $\pi$ energy states of quantum walks. 
It is of great interests to study if other types of topological phenomena unique to driven systems can be realized 
in quantum walks. 

\section{Conclusion and open questions}  \label{sec:conclusion}
In this review article, we studied topological phases appearing in quantum walks. 
After the introduction of quantum walks, 
we provided a thorough explanation of topological nature of quantum walks. 
We first associated the quantum walks with winding numbers, and gave an intuitive argument for 
the existence of bound states across the boundary of the regions that 
belong to different topological phases. 
We argued for the topological protection
of bound states in two different point of view; one from the spectrum (gap) in the system and another from
topological invariant associated with the bound states. 
These physics are illustrated through the explicit example of quantum walks in 
one and two dimensions. We also explicitly demonstrated the existence of topological phenomena unique to
periodically driven systems in one and two dimensions. 

There are many open questions that one can study in the field of quantum walks. 
For example, there is not yet an example of three dimensional quantum walks that realizes non-trivial 
topological phase. A simple example of such quantum walks are of interests, considering the excitement in the 
field of three dimensional topological insulators\cite{Hasan2010,Qi2011}. Moreover, three dimensional quantum walks 
with spin $1/2$ has the possibility to realize Hopf-insulator 
first proposed by [\onlinecite{Moore2008}]. Since the realization of 
this topological phase is very difficult in condensed matter materials, it is of great interests to explore the 
possibility of realizations in artificial systems such as quantum walks. 

Other open direction is provided by quantum walks in different geometries, such as hexagonal lattice. 
Since hexagonal lattice has three neighbors, study of hexagonal lattice quantum walk with spin $1$ might provide
interesting platform to explore unique quantum phenomena. 

Less concretely, the study of quantum walks with a few to many particles with strong correlation would be interesting 
to investigate. In particular, in the presence of frustrated hopping, there may be unique quantum phenomena such as 
the formation of spin-liquid phase. 

The author would like to thank a numerous insightful discussions with Mark S. Rudner, Erez Berg, Yutaka Shikano, and Eugene Demler. 
 
\appendix

\section{Asymptotic distribution of quantum walk} \label{appendix:asymptotic}
In this section, we derive the intuitive result \eqnref{distribution} which gives the 
asymptotic distribution of quantum walks.  Here we consider a quantum walk initially prepared at site $x=0$
with initial state $\ket{s}$ such that $\ket{i} = \ket{x=0} \otimes \ket{s}$. 
The evolution of a particle after each step is dictated by the effective Hamiltonian 
given by $H_{\textrm{eff}} = \int dk E(k) \vec{n}(k) \cdot \sigma \otimes \ket{k}\bra{k} $ as in \eqnref{Hk}. 

Since the particle is propagating under the non-interacting Hamiltonian $H_{\textrm{eff}}$, it is natural to
expect that the particle propagates in a ballistic fashion. Thus, the particle distribution has a well-defined form 
in terms of the variable $X = x/N$ in the asymptotic limit. The distribution of $X$, $P(X)$, can be computed through 
\begin{eqnarray}
\label{characteristic_function}
P(X) &=& \braket{ \delta(\hat{x}/N - X)}  \nonumber \\
&=& \braket{ \int^{\infty}_{-\infty} ds \, e^{is (X - \hat{x}/N)} } \nonumber \\
&=&\int^{\infty}_{-\infty} ds \, e^{isX}  \braket{  e^{-is \hat{x}/N}}
\end{eqnarray}
Thus we aim to obtain the expectation of so-called characteristic function $e^{-is \hat{x}/N}$ after $N$th steps 
as $N \rightarrow \infty$. 

We first note that $\hat{x} = \sum_{j} j \ket{j} \bra{j}$, and thus 
\begin{eqnarray*}
e^{-is \hat{x}/N} &=& 1 + \sum_{j} (- is) \frac{j}{N} \ket{j} \bra{j} + (- is)^2 \frac{(j/N)^2}{2!} \ket{j} \bra{j} + \cdots \\
&=& 1 + \sum_{j} \left( e^{-is j/N} -1 \right)  \ket{j} \bra{j} \\
 &=&  \sum_{j}  e^{-is j/N}  \ket{j} \bra{j} \\
  &=&  \int dk \,   \ket{k+ s/N} \bra{k}
 \end{eqnarray*}
 Now we evaluate $ \braket{  e^{-is \hat{x}/N}} = 
 \bra{i} e^{iH_{\textrm{eff}} N} e^{-is \hat{x}/N} e^{-iH_{\textrm{eff}} N} \ket{i} $. 
 Since $H_{\textrm{eff}}$ is diagonal in quasi-momentum space, the evaluation is straightforward. First of all, 
 \begin{eqnarray*}
 && e^{iH_{\textrm{eff}} N} e^{-is \hat{x}/N} e^{-iH_{\textrm{eff}} N} \\ 
 &=& 
  \int dk \,   \ket{k+ s/N} \bra{k}  \otimes e^{i N E(k+s/N) \vec{n}(k+s/N) \cdot \sigma} \\
 && \times e^{- i N E(k) \vec{n}(k) \cdot \sigma} \\
&= &   \int dk \,   \ket{k+ s/N} \bra{k} \otimes  e^{i s v_{k}  \vec{n}(k) \cdot \sigma} 
   \end{eqnarray*}
   In the last line, we took the expression in the lowest order in $s/N$. This can be confirmed through 
   the expansion
    $\exp\left\{ N E(k+s/N) \vec{n}(k+s/N) \cdot \sigma \right\} = 
   \cos \left\{ N E(k+s/N)  \right\} + i \sin  \left\{ N E(k+s/N)  \right\} \vec{n}(k+s/N) \cdot \sigma \approx 
   \cos \left( N E(k) + sv_{k}  \right) + i \sin \left( N E(k) + sv_{k}  \right)  \vec{n}(k) \cdot \sigma$ to the lowest
   order in $\frac{s}{N}$. 

It is now straightforward to evaluate the expectation value of above expression in the initial state
$\ket{i} = \int^{\pi}_{-\pi}  \frac{dk}{\sqrt{2\pi}} \ket{k} \otimes \ket{s}$.  Using the expression 
\eqnref{characteristic_function}, we obtain the final expression 
\begin{eqnarray*}
P(X) &=& \int^{\pi}_{-\pi} \frac{dk}{2\pi} \frac{1}{2} \left( 1+ \braket{\vec{n}(k) \cdot \sigma} \right) \delta(v_{k} - X) \nonumber \\ 
&&+ \frac{1}{2} \left( 1- \braket{\vec{n}(k) \cdot \sigma} \right) \delta(v_{k} + X) 
\label{distribution} 
\end{eqnarray*}


\section{Sublattice (chiral) symmetry of inhomogeneous quantum walk} \label{appendix:sublatice_symmery}
In this section, we give the explicit proof of sublattice (chiral) symmetry for inhomogeneous split-step quantum walks. 
In \eqnref{chiralsymmetry}, we defined the sublattice (chiral) symmetry in terms of Hamiltonian. This definition 
directly translates to the sublattice (chiral) symmetry on the evolution operator after one period $U$ as 
\begin{equation}
\Gamma^{-1} U \Gamma = U^{\dagger}
\end{equation} 
We have shown in \secref{sec:splitstep} that the (homogeneous) split-step quantum walk,
$U= T_{\downarrow}R_{y}(\theta_{2}) T_{\uparrow}R_{y}(\theta_{1})$ possesses the symmetry with the operator
$\Gamma_{\theta_{1}}{=}e^{-i\pi \vec{A} \cdot \vec{\sigma}/2}$ 
where $\vec{A} = (\cos\theta_{1}/2, 0, -\sin\theta_{1}/2)$. Here we write the subscript $\theta_{1}$ 
on the symmetry operator to emphasize the dependence on $\theta_{1}$. 

Here  $\Gamma_{\theta_{1}}$ is a local operator and thus, we expect that the chiral symmetry is preserved 
even if $\theta_{2}$ becomes inhomogeneous in space.

In order to explicitly check this, we expand the evolution operator 
$U= T_{\downarrow}R_{y}(\theta_{2}) T_{\uparrow}R_{y}(\theta_{1})$ in the position basis, where 
we take the general case that $\theta_{2}$ depends on space in an arbitrary fashion. For example, 
$T_{\uparrow}{=}\sum_{x} (1+\sigma_{z})/2 \ket{x+1}\bra{x}+ (1-\sigma_{z})/2 \ket{x}\bra{x}$.
After the expansion, one obtains 
\begin{eqnarray*}
U &{=}&  \sum_{x} \frac{1+\sigma_{z}}{2} R_{y}(\theta_{2}(x+1))  \frac{1+\sigma_{z}}{2} R_{y}(\theta_{1}) \otimes  \ket{x+1}\bra{x} \\ 
&&+ \frac{1-\sigma_{z}}{2} R_{y}(\theta_{2}(x+1))  \frac{1-\sigma_{z}}{2} R_{y}(\theta_{1})  \otimes \ket{x}\bra{x+1}  \\ 
&& + \left( \frac{1-\sigma_{z}}{2} R_{y}(\theta_{2}(x+1))  \frac{1+\sigma_{z}}{2} R_{y}(\theta_{1})  \right. \\ 
&& \left. +\frac{1+\sigma_{z}}{2} R_{y}(\theta_{2}(x))  \frac{1-\sigma_{z}}{2} R_{y}(\theta_{1}) \right)  \ket{x}\bra{x} 
\end{eqnarray*}
Now the sublattice (chiral) symmetry condition 
$\left( \Gamma'_{\theta_1} \right)^{-1} U \Gamma_{\theta_1} = U^{\dagger}$
can be checked by comparing both sides of the equation for 
each position operators of the form $\ket{x}\bra{x+\alpha}$ with $\alpha=-1,0,1$. 
For example, comparing the both sides of the equation for the coefficients of $\ket{x+1}\bra{x}$,
sublattice (chiral) symmetry requires that
\begin{eqnarray*}
& & \left( \Gamma_{\theta_1} \right)^{-1} \frac{1+\sigma_{z}}{2} R_{y}(\theta_{2}(x+1))  \frac{1+\sigma_{z}}{2} R_{y}(\theta_{1})\Gamma_{\theta_1}  \\
& &  \stackrel{?}{=} R_{y}^{-1}(\theta_{1}) \frac{1-\sigma_{z}}{2} R_{y}^{-1}(\theta_{2}(x+1))  \frac{1-\sigma_{z}}{2}
\end{eqnarray*} 
It is straightforward to check that this equality indeed holds for any rotation $\theta_{1}$ and $\theta_{2}(x+1)$. 
Repeating such process for the coefficients of $\ket{x}\bra{x+1}$ and $\ket{x}\bra{x}$, one confirms the existence of sublattice (chiral) symmetry for inhomogeneous quantum walks. 

Notice that the split step quantum walk with $\theta_{1} =0$ is effectively the conventional quantum walk
described in \secref{sec:intro}. The quantum walk becomes $U = T_{\downarrow} R_{y}(\theta_{2}) T_{\uparrow}$,
which is unitarily related to the conventional quantum walk $U_{\textrm{con}} = T R_{y}(\theta_{2}) = T_{\uparrow} T_{\downarrow} R_{y}(\theta_{2})$ by the shift of time. Therefore, the explanation above also provides the proof that
the disordered conventional quantum walk where the rotation angle at each site is random possesses sublattice (chiral) symmetry. 

\section{Analytic solution of the bound state for quantum walks with reflecting boundary condition} \label{appendix:bound_state}
In this section, we give the analytical solution of bound states for quantum walks with reflecting boundary condition 
studied in \secref{sec:reflecting}.
Given the quantum walk with reflecting boundary condition whose evolution operator is 
$U_{x \leq 0}$ in \secref{sec:reflecting}, we look for the bound states near $x=0$. 
Generally, such bound state can be written as 
\begin{eqnarray} \label{eigen}
\ket{\psi_{b}} = \sum_{j \le 0 } \left( c_{j,\downarrow} \ket{\downarrow}  + c_{j,\uparrow} \ket{\uparrow} \right) \ket{j} 
\end{eqnarray}

The approach we take is to directly solve the eigenvalue problem 
\begin{equation}
U_{x \leq 0} \ket{\psi_{b}} = e^{-iE_{b}}  \ket{\psi_{b}}
\end{equation}
where $E_{b}$ is the quasi-energy of the bound state. Comparison of the two sides of 
the equation above together with the normalizability of the bound state wavefunction allows the 
solution of the problem. 

The left-hand side of the equation gives 
\begin{eqnarray*}
&& U_{x \leq 0} \ket{\psi_{b}} \\
& =&  \sum_{j \le 0 } \tilde{c}_{j,\downarrow} \ket{j-1, \downarrow}  + 
\sum_{j \le -1} \tilde{c}_{j,\uparrow} \ket{j+1, \uparrow}  + e^{i\phi} \tilde{c}_{0,\uparrow} \ket{0,\downarrow} \\
& = & \sum_{j \le -1 } \tilde{c}_{j+1,\downarrow} \ket{j, \downarrow} +
\sum_{j \le 0} \tilde{c}_{j-1,\uparrow}  \ket{j, \uparrow}  + e^{i\phi} \tilde{c}_{0,\uparrow} \ket{0,\downarrow} 
\end{eqnarray*}
where the tilde coefficients $ \tilde{c}_{j,\uparrow, \downarrow}$ are related to 
the original coefficients $ c_{j,\uparrow, \downarrow}$ through the rotation $R_{y}(\theta)$ as 
\begin{eqnarray*}
\left( \begin{array}{c} \tilde{c}_{j, \uparrow} \\ \tilde{c}_{j, \downarrow} \end{array}  \right)  
& = &   R_{y}(\theta)
\left( \begin{array}{c} c_{j, \uparrow} \\ c_{j, \downarrow} \end{array}  \right) \\
&=&  \left( \begin{array}{cc} \cos(\theta/2) & -\sin(\theta/2)  \\ \sin(\theta/2) & \cos(\theta/2)  \end{array} \right)  
\left( \begin{array}{c} c_{j, \uparrow} \\ c_{j, \downarrow} \end{array}  \right)  \\
\end{eqnarray*}

Comparison of the equation above with the right hand side of \eqnref{eigen} immediately gives
\begin{eqnarray}
 e^{-iE_{b}}  c_{j,\downarrow} &=&  \tilde{c}_{j+1,\downarrow} \quad  j \le -1 \nonumber \\
e^{-iE_{b}}  c_{j,\uparrow} &=&  \tilde{c}_{j-1,\uparrow} \quad  j \le 0 \nonumber  \\
 e^{-iE_{b}}  c_{0,\downarrow} &=& e^{i\phi} \tilde{c}_{0,\uparrow}  \label{zeroconstraint}
\end{eqnarray}

In matrix form, the first two equations can be rewritten as
 \begin{widetext}
\begin{eqnarray}
&&  \left( \begin{array}{cc} 0 & e^{-iE_{b}} \\  \cos(\theta/2) &  - \sin(\theta/2)  \end{array} \right)  
   \left( \begin{array}{c} c_{j,\uparrow} \\ c_{j,\downarrow}   \end{array} \right)   =  
 \left( \begin{array}{cc} \sin(\theta/2) &  \cos(\theta/2) \\ e^{-iE_{b}}  & 0  \end{array} \right)   
 \left( \begin{array}{c} c_{j+1,\uparrow} \\ c_{j+1,\downarrow}   \end{array} \right) \quad \textrm{for $j \leq -1$} \nonumber \\
& \rightarrow &  \left( \begin{array}{c} c_{j,\uparrow} \\ c_{j,\downarrow}   \end{array} \right)  = 
e^{iE_{b}}
 \left( \begin{array}{cc} \frac{ \sin^2(\theta/2)}{\cos(\theta/2)} + 
 \frac{1}{\cos(\theta/2)} e^{-2iE_{b}} &  \sin(\theta/2)  \\  \sin(\theta/2)  & \cos(\theta/2)  \end{array} \right)   
 \left( \begin{array}{c} c_{j+1,\uparrow} \\ c_{j+1,\downarrow}  \end{array}  \right)  \quad \textrm{for $j \leq -1$} \label{recursion}
 \end{eqnarray}
 \end{widetext}
This last equation is a recursive equation that relates the coefficients at site $j+1$ to site $j$. 
We denote the matrix that relates them as ${\bf K}$, which is a matrix that appears on the right hand side 
of \eqnref{recursion}. 

The behavior of wavefunction in the limit of $x \rightarrow -\infty$ is determined by 
the eigenvalues of the matrix ${\bf K}$. They are given by 
 \begin{eqnarray*}
 {\bf K} &=& 
    \left( \begin{array}{cc} \vec{v}_{+} & \vec{v}_{-}  \end{array}  \right) 
    \left( \begin{array}{cc} \lambda_{+} & 0 \\  0 & \lambda_{-}  \end{array}  \right)
  \left( \begin{array}{c} \vec{v}_{+}^{T} \\ \vec{v}_{-}^{T}   \end{array}  \right) \\  
  \vec{v}_{\pm} & = & \frac{1}{N_{\pm}} \left( \begin{array}{c} \frac{ e^{-2iE_{b}} - \cos(\theta) \pm e^{-iE_{b}} \sqrt{
  e^{-2iE_{b}}+ e^{2iE_{b}} -2 \cos(\theta)} } {\sin(\theta)} \\ 1 \end{array}  \right) \\
 \lambda_{\pm} &=& \frac{  \cos(E_{b}) \pm  \sqrt{
\cos^2(E_{b}) -  \cos^2(\theta/2)}  }{ \cos(\theta/2)} 
 \end{eqnarray*}
 where $N_{\pm}$ in the expression of $ \vec{v}_{\pm} $ are the normalization factors. 
 
Then, the amplitude of bound states wavefunction at site $-j$ is given by
  \begin{eqnarray*}
   \left( \begin{array}{c} c_{-j,\uparrow} \\ c_{-j,\downarrow}   \end{array} \right)  & = & 
Q \left( \begin{array}{cc} \lambda_{+}^{j} & 0 \\ 0 & \lambda_{-}^j  \end{array} \right)   Q^{-1}
 \left( \begin{array}{c} c_{0,\uparrow} \\ c_{0,\downarrow}  \end{array}  \right) \quad j \le -1
 \end{eqnarray*}
 where $Q =  \left( \begin{array}{cc} \vec{v}_{+} & \vec{v}_{-}  \end{array}  \right) $. 
 One crucial observation is $\lambda_{+} \lambda_{-} =1$, and therefore
 $|\lambda_{+}| \leq 1$ when $\cos(E_{b}) \leq 0$ and  $|\lambda_{-}| \leq 1$ when $\cos(E_{b}) \geq 0$. 
  The normalizability of the bound state wavefunction requires that the amplitude
 $( c_{0,\uparrow}, c_{0,\downarrow} )^{T}$ is proportional to $\vec{v}_{+} (\vec{v}_{-})$
 when  $\cos(E_{b}) \leq 0 (\cos(E_{b}) \geq 0)$. No normalizable bound state wavefunction 
 exists when $|\lambda_{+}| = |\lambda_{-}| =1$, or $\cos^2(E_{b}) -  \cos^2(\theta/2) <0$. 
 
 Additional constraint on the amplitudes  $c_{0,\uparrow}, c_{0,\downarrow}$ comes 
 from the equation \eqnref{zeroconstraint}, namely,  $e^{-iE_{b}}  c_{0,\downarrow} = e^{i\phi} \tilde{c}_{0,\uparrow}$.
 Solving these two conditions give us 
  \begin{eqnarray*}
\sin(\theta/2) e^{-i\phi} = -i \sin(E_{b}) \mp \sqrt{ \cos^2(E_{b}) - \cos^2(\theta/2)}
\end{eqnarray*}
 where minus sign is for $v_{-}$ or when $\cos(E_{b}) \geq 0$ and plus sign is 
 for $v_{+}$ or when $\cos(E_{b}) \leq 0$. 
  
 When $\phi =0$ and $0 < \theta < 2\pi$, the solution exists 
 at energy $E_{b}=\pi$ with $( c_{0,\uparrow}, c_{0,\downarrow} )^{T} \propto v_{+}$.
 For $\phi =0$ and $2 \pi < \theta < 4 \pi$, the bound state energy is 
$E_{b}=0$ and $( c_{0,\uparrow}, c_{0,\downarrow} )^{T} \propto v_{-}$.
 
 On the other hand, when $\phi=\pi$ and $0 < \theta < 2\pi$, the 
 bound state energy is $E_{b}=0$ with $( c_{0,\uparrow}, c_{0,\downarrow} )^{T} \propto v_{-}$ ,
 whereas when $\phi=\pi$ and $2 \pi < \theta < 4\pi$, the 
 bound state energy is $E_{b}=\pi$ with $( c_{0,\uparrow}, c_{0,\downarrow} )^{T} \propto v_{+}$. 
 
The bound state wave function found above decays on the 
length scale of $\sim 1/|\log(\lambda_{-})| = \frac{1}{|\log(1- |\sin(\theta/2)|) - \log(\cos(\theta/2))|}$. Thus 
the extent of bound state approaches $\infty$ as $\theta \rightarrow 0, 2\pi$. On the other hand,
the bound state becomes most localized when $\theta =\pi$. 

\section{Spectrum of two dimensional quantum walk} \label{appendix:2D_spectrum}
Here we give the details of how to compute the spectrum of two dimensional quantum walk introduced in \secref{sec:2D_quantum_walk}. 
The method introduced here is general and can be easily extended to other protocols of quantum walks 
in, say, higher dimensions. 

The evolution operator of one step for the two dimensional quantum walk can be written, in the quasi-momentum space,
as 
\begin{eqnarray*}
U(k_{x}, k_{y}) = e^{ik_{x} \sigma_{z}} e^{-i\theta_{1} \sigma_{y}/2} e^{ik_{y} \sigma_{z}} e^{-i\theta_{2} \sigma_{y}/2} \\
\times e^{i(k_{x} + k_{y}) \sigma_{z}} e^{-i\theta_{1} \sigma_{y}/2} 
\end{eqnarray*}

Most general form of the effective Hamiltonian resulting from spin $1/2$ system is given by
\begin{equation}
H_{\textrm{eff}}(\vec{k}) = E_{0}(\vec{k}) + E(\vec{k}) \vec{n} (\vec{k}) \cdot \sigma
\end{equation} 
This is true because a generator of two by two unitary matrix is ${\bf 1}$ and Pauli matrices $\vec{\sigma}$. 

Then the spectrum can be identified by considering the trace of evolution operator because
\begin{eqnarray*}
&& \textrm{Tr} ( U(k_{x}, k_{y}) ) \equiv \textrm{Tr} \left( \exp (-iH_{\textrm{eff}}(\vec{k})  ) \right) \\
& =&\textrm{Tr} \left\{  e^{-i E_{0}(\vec{k}) } \left( \cos \left( E(\vec{k}) \right) - i \sin \left( E(\vec{k}) \right) \vec{n} (\vec{k}) \cdot \sigma \right) \right\} \\
& =& 2  e^{-i E_{0}(\vec{k}) }  \cos \left( E(\vec{k}) \right)
\end{eqnarray*}

The explicit evaluation of the trace of $U(k_{x}, k_{y})$ shows that $ E_{0}(\vec{k}) =0 $ for all $\vec{k}$ 
and 
\begin{eqnarray*}
 \cos \left( E(\vec{k}) \right) &=& \cos( k_{x}) \cos( k_{x}+ 2k_{y}) \cos(\theta_{1}) \cos(\theta_{2}/2) \\ 
&&- \sin( k_{x}) \sin( k_{x}+ 2k_{y}) \cos(\theta_{2}/2)  \\
&& - \cos^2( k_{x}) \sin( \theta_{1}) \sin(\theta_{2}/2) 
\end{eqnarray*}

\section{gapless phase of two dimensional quantum walk} \label{appendix:2D_gapless}
In this section, we detail the calculation to obtain the line of gapless phase in the phase diagram of \fref{fig:2D_phase_diagram} a). 
The two bands of the system closes the gap when the two eigenvalues of Hamiltonian 
\begin{equation}
H_{\textrm{eff}}(\vec{k}) = E(\vec{k}) \vec{n} (\vec{k}) \cdot \sigma
\end{equation} 
becomes degenerate. Since the eigenvalues of $\vec{n} (\vec{k}) \cdot \sigma$ is $\pm 1$, the 
quasi-energy of the states become degenerate if $E(\vec{k}) = - E(\vec{k})$, which happens if 
$E(\vec{k}) = 0$ or $\pi$ due to the periodicity of the quasi-energy. 

The gapless phase occurs at the values of $\theta_{1}$ and $\theta_{2}$ such that the equation 
\eqnref{2D_energy} has a solution of $E(\vec{k}) = 0$ or $\pi$ for some value of $k_{x}$ and $k_{y}$. 
On the other hand, at $E(\vec{k}) = 0$ or $\pi$, 
the evolution operator $U(k_{x}, k_{y}) $ takes the value $1$ or $-1$, respectively.

A simple way to obtain such values of $\theta_{1}$ and $\theta_{2}$ is to look at $(1,1)$ component
of the evolution operator $a_{11} = U(k_{x}, k_{y})[1,1] $. 
First we study the lines of gapless phases where the gap is closed at the quasi-energy $0$. 
Then the equation $1 = a_{11}$ gives the condition 
\begin{eqnarray}
&&  1= e^{ik_{x}} \left\{ i \sin(k_{x} + 2 k_{y}) \cos(\theta_{2}/2)  \right. \nonumber \\
&&   + \cos(k_{x} + 2k_{y}) \cos \theta_{1} \cos(\theta_{2}/2) \nonumber \\
&& \left.  - \cos k_{x} \sin\theta_{1} \sin(\theta_{2}/2) \right\} \label{cond}
 \end{eqnarray}
 Note that $1$ is the maximum magnitude that RHS of the above equation attains for any values 
 of $k_{x}, k_{y}, \theta_{1}$ and $\theta_{2}$. Therefore, 
 we can simply maximize the RHS in terms of the variables $k_{x}$ and $k_{y}$ or alternatively
 the variables $k_{x} + 2k_{y}$ and $k_{x}$. The argument separates few cases. 
 
 If  $\sin\theta_{1} \sin(\theta_{2}/2)$ is non-zero, then $\cos k_{x} = \pm 1$. If we take the two orthogonal 
 variables $k_{1} = k_{x} + 2 k_{y}$ and $k_{2} = 2k_{x} - k_{y}$, then the first two terms of the RHS of 
 \eqnref{cond} is only a function of $k_{1}$. By differentiating the absolute square of RHS in terms of $k_{2}$, 
 one can show that the extremum of this value is taken when  $\cos k_{x} = \pm 1$. 
 
 Suppose $\cos k_{x} = 1$. Then $e^{ik_{x}} =1$ and the equation \eqnref{cond} requires 
 the first term $\sin(k_{x} + 2 k_{y}) \cos(\theta_{2}/2)$ to be zero. If we take $\sin(k_{x} + 2 k_{y})$ to be 
 zero, then $\cos(k_{x} + 2 k_{y}) = \pm 1$, and the condition is reduced to 
 $ 1 = \cos(\theta_{1} + \theta_{2}/2)$ for plus sign and  $ -1 = \cos(\theta_{1} - \theta_{2}/2)$ for minus sign. 
 Thus the gapless phase exist whenever $\theta_{1} + \theta_{2}/2 = 2\pi n$ and $\theta_{1} - \theta_{2}/2 = 2\pi n + \pi$. 
 If $\cos(\theta_{2}/2) =0$, then the equation is solved only at discrete points of $\theta_{1} = \pi/2 + 2\pi n$ 
 with $\theta_{2} =  3\pi + 4\pi n$ or $\theta_{1} = 3\pi/2 + 2\pi n$ with $\theta_{2} =  \pi + 4\pi n$. These cases 
 are already included in the condition above. 
 
 Similarly consider $\cos k_{x} = -1$. Then $e^{ik_{x}} =-1$ and the equation \eqnref{cond} again requires 
 the first term $\sin(k_{x} + 2 k_{y}) \cos(\theta_{2}/2)$ to be zero.  If we take $\sin(k_{x} + 2 k_{y})$ to be 
 zero, then $\cos(k_{x} + 2 k_{y}) = \pm 1$, and the condition is reduced to 
 $ -1 = \cos(\theta_{1} - \theta_{2}/2)$ for plus sign and  $ 1 = \cos(\theta_{1} + \theta_{2}/2)$ for minus sign. 
 Thus the gapless phase exist whenever $\theta_{1} - \theta_{2}/2 = \pi + 2\pi n$ and $\theta_{1} + \theta_{2}/2 = 2\pi n$. 
 Thus these conditions give exactly the same gapless phases as for  $\cos k_{x} = 1$. 
 
 Now suppose that $\sin\theta_{1} \sin(\theta_{2}/2)$ is zero. A new condition appears when $ \sin(\theta_{2}/2) =0$. 
 Then $\cos(\theta_{2}/2) = \pm 1$. But now, we can satisfy the condition \eqnref{cond} by setting 
 $\cos(k_{x} + 2k_{y}) = 0$. Since this does not require for $\theta_{1}$ to take any particular value, we conclude that
 gap closes for the line of $\theta_{2} = n \pi$. 
 
 A similar consideration for quasi-energy $E=\pi$ gives the condition that the gap closes at $E=\pi$ for 
 $\theta_{1} + \theta_{2}/2 = 2\pi n + \pi$, $\theta_{1} - \theta_{2}/2 = 2\pi n $ and $\theta_{2} = n \pi$. 
 These results lead to the gapless phases plotted in \fref{fig:2D_phase_diagram} a).

\bibliography{qwreview}{}

\begin{thebibliography}{10}
\bibitem{dqw}
Throughout the paper, we consider discrete quantum walks, so we interchangeably
  use the word "discrete quantum walk" and "quantum walk" when confusion does
  not arise.
\bibitem{footnote1}
These two words are used interchangeably in the literature. The word,
  "sublattice symmetry" is intuitive. When one has a Hamiltonian, for example,
  with sublattice structure with sublattice $A$ and $B$ such that the
  Hamiltonian is only hopping from sites of $A$ to sites of $B$, then
  Hamiltonian can be made minus of itself by multiplying all the creation and
  annihilation operators of sublattice $A$. This operation is unitary, and
  thus, the Hamiltonian possesses sublattice symmetry. The word, "chiral
  symmetry," originates from the high energy physics, where it relates the two
  handedness of Dirac fields. Because the word "chiral" has the meaning of
  "handedness," the world "chiral symmetry" is also used in solid state physics
  for some crystal symmetries, and care has to be taken to use the word.
\bibitem{footnote3}
In the demonstration, the rotation angle $\theta_{2}$ is constant and the first
  rotation angle $\theta_{1}$ is varied in space. As we have explained at the
  end of \secref{sec:splitstep}, such quantum walk also possess sublattice
  (chiral) symmetry and thus the essential physics remains the same.
\end{thebibliography}
\bibliographystyle{h-physrev}
\end{document}